\documentclass[twocolumn]{aastex631}

\usepackage[normalem]{ulem}
\usepackage{comment}

\shorttitle{Be star demographics}
\shortauthors{Figueiredo et al.}
\graphicspath{{./}{figures/}}

\begin{document}

\title{Be star demographics: a comprehensive study of thousands of lightcurves in the Magellanic Clouds}

\author[0009-0007-1625-8937]{André L. Figueiredo}
\affiliation{Instituto de Astronomia, Geof{\'i}sica e Ci{\^e}ncias Atmosf{\'e}ricas, Universidade de S{\~a}o Paulo,\\Rua do Mat{\~a}o 1226, Cidade Universit{\'a}ria, B-05508-900 S{\~a}o Paulo, SP, Brazil}

\author[0000-0002-9369-574X]{Alex C. Carciofi}
\affiliation{Instituto de Astronomia, Geof{\'i}sica e Ci{\^e}ncias Atmosf{\'e}ricas, Universidade de S{\~a}o Paulo,\\Rua do Mat{\~a}o 1226, Cidade Universit{\'a}ria, B-05508-900 S{\~a}o Paulo, SP, Brazil}

\author[0000-0002-2919-6786]{Jonathan Labadie-Bartz}
\affiliation{LIRA, Observatoire de Paris, Université PSL, CNRS, Sorbonne Université\\Universit\'e Paris Cit\'e, CY Cergy Paris Universit\'e, 92190 Meudon, France}

\author[0009-0004-4232-1651]{Matheus L. Pinho}
\affiliation{Instituto de Astronomia, Geof{\'i}sica e Ci{\^e}ncias Atmosf{\'e}ricas, Universidade de S{\~a}o Paulo,\\Rua do Mat{\~a}o 1226, Cidade Universit{\'a}ria, B-05508-900 S{\~a}o Paulo, SP, Brazil}

\author[0000-0001-5563-6629]{Tajan H. de Amorin}
\affiliation{Instituto de Astronomia, Geof{\'i}sica e Ci{\^e}ncias Atmosf{\'e}ricas, Universidade de S{\~a}o Paulo,\\Rua do Mat{\~a}o 1226, Cidade Universit{\'a}ria, B-05508-900 S{\~a}o Paulo, SP, Brazil}

\author[0000-0002-4808-7796]{Pedro Ticiani dos Santos}
\affiliation{Instituto de Astronomia, Geof{\'i}sica e Ci{\^e}ncias Atmosf{\'e}ricas, Universidade de S{\~a}o Paulo,\\Rua do Mat{\~a}o 1226, Cidade Universit{\'a}ria, B-05508-900 S{\~a}o Paulo, SP, Brazil}

\author[0000-0002-7777-0842]{Igor Soszy{\'n}ski}
\affiliation{Astronomical Observatory, University of Warsaw, Al. Ujazdowskie 4, 00-478\\Warszawa, Poland}

\author[0000-0001-5207-5619]{Andrzej Udalski}
\affiliation{Astronomical Observatory, University of Warsaw, Al. Ujazdowskie 4, 00-478\\Warszawa, Poland}

\correspondingauthor{André L. Figueiredo}
\email{andre.luiz.figueiredo@usp.br}

\begin{abstract}

Multi-color OGLE survey light curves of about 20 years duration are analyzed for about 3000 classical Be stars in the Large and Small Magellanic Clouds (LMC, SMC) in order to study the properties and variability.
Each light curve was manually analyzed to distinguish between different scenarios, such as photospheric baseline levels, and disk build-up and dissipation phases.
This analysis was aided by dynamical disk models and photospheric models to coarsely determine inclination angle and mass.
Measured quantities such as the fraction of time spent actively ejecting mass (the duty cycle), the fraction of time spent with a detectable disk (the disk duty cycle), the build-up and dissipation time of isolated disk events, and the number of mass outbursts per year allow us to characterize and compare the behavior of the two populations.
There is a wide spread in the duty cycle, with median values of 0.44 (LMC) and 0.60 (SMC).
The disk duty cycle is high for both populations, with median values of 0.99 (LMC) and 1.0 (SMC), indicating that disks are almost always present for these stars.
The occurrence rate of outbursts ranges from zero to about two per year, with median values of 0.31 (LMC) and 0.26 (SMC).
There are strong statistical differences in the behavior of the LMC and SMC populations, with the lower metallicity stars being more active in terms of their duty cycle and disk duty cycle, and with less frequent but longer lasting outbursts.

\end{abstract}


\section{Introduction} \label{sec:intro}
\label{sec:introduction}

A star can lose mass through several mechanisms, such as stellar winds, in binary systems where the star fills its Roche lobe and transfers mass to its companion, or ejecting its outer layers in advanced stages of evolution, along with other processes.
Studying these mechanisms offers important insights on stellar structure and evolution, as the expelled material can dramatically affect local conditions.

A peculiar way some massive and intermediate-mass stars can lose both mass and angular momentum is by ejecting a disk of material, 
a {process that, in its individuality, is unique to Be stars.}
This is possible due to two characteristics all Be stars share: high rotation rates coupled with non-radial pulsations.
The typical rotation rate, expressed as the ratio between the equatorial rotation speed to the orbital speed at the equatorial radius, of Galactic stars is $W = V_{\rm rot} / V_{\rm orb} \approx 0.8$ \citep{Rivinius2013A&ARv..21...69R}, which makes expelling material much easier.
{The best candidate to provide the extra energy to put material into orbit is non-radial pulsations \citep{Owocki_2006ASPC..355..219O,baade_2016A&A...588A..56B}.}
Once material is ejected from the stellar surface, its dynamics are governed by viscosity.
Arguably, variability {is one of the most prominent characteristics} of Be stars, {arising from the complex interplay of pulsation, rotation, mass ejection, and circumstellar changes, often resulting in apparently erratic behavior.}
These fluctuations can occur on timescales ranging from hours to decades, providing a unique laboratory for studying viscous dynamics and other processes.

Except for a minority of cases of well-behaved Be stars, which display remarkably stable disks \citep[a notable case being $\beta$~CMi,][]{klement_2015}, studying and reproducing the variability of Be stars and their disks presents a significant challenge that only recently has been addressed with some degree of success.
Modern developments began with \citet{Lee_1991}, who proposed the Viscous Decretion Disk (VDD) model.
This model was inspired by the theory of accretion disks around black holes, as outlined by \cite{Shakura_1973A&A....24..337S}, with the key distinction that both matter and angular momentum flow outward in Be star disks.
Further contributions were made by \cite{Okazaki_2001PASJ...53..119O}, \cite{bjorkman_e_carciofi_2005} and \cite{Carciofi_2008ApJ...684.1374C}, among many other authors.
The model suggests that, once material is placed into orbit around the stellar equator, viscosity plays a significant role in governing the dynamics of the circumstellar material, allowing the disk to spread outward through viscous torques.
The VDD model is regarded as the most promising theory to explain observations, including data obtained through various techniques.
It has been used to model individual stars \citep[e.g.,][]{Carciofi_2006ApJ...652.1617C,klement_2015,Marr_2021ApJ...912...76M} as well as samples of stars 
\citep[e.g.,][]{Touhami_2011ApJ...729...17T,Rimulo_2018MNRAS.476.3555R,Vieira_2017MNRAS.464.3071V}, to name a few examples.

It is important to note that other mechanisms, beyond viscous driving, may also be at play in these systems, and their effects require further understanding.
One such example is disk ablation, which has been explored by \citet{Kee_2016MNRAS.458.2323K}, although a detailed comparison with observations is still lacking.
Furthermore, modeling the temporal evolution of well-documented Be disks, as performed by \citet{Marr_2021ApJ...912...76M} for 66\,Oph, \citet{2021ApJ...909..149G} for $\omega$\,CMa, 
and more recently, \citet{2024MNRAS.533.2867G} for MT91-213, all suggest that the predictions of the VDD theory fall short in reproducing the visible and/or polarization light curves 
\citep[which originate in the inner disk, e.g.,][]{Carciofi_2011IAUS..272..325C}, as well as the emission line properties (which can be traced to a much larger emission volume). 
The reasons for these shortcomings remain to be investigated.

Given the extensive testing and validation that the VDD theory has undergone, it can be considered a reliable framework for interpreting the physical mechanisms driving the variability of Be star disks, despite its current limitations.
However, the main challenge in understanding the broader picture is the erratic nature of the Be phenomenon and the quite diverse behavior displayed among Be stars, even among those of similar mass, age and metallicity.
One way to move forward in this regard is through population studies.
For instance, the study of Be populations in clusters can represent a key advance in better understanding their evolution.
The measurement of the fraction of Be stars over the B population in clusters of different ages can help us to estimate the lower limit of how common the Be phenomenon is \citep[for instance,][that estimated be star fractions for clusters in the Magellanic clouds]{Martayan2006A&A...452..273M, Martayan_2007A&A...462..683M}.
Another approach involves a systematic analysis of a large sample of Be stars observed over an extended period, which is the main focus of this paper. We analyze 2144 light curves of Be stars in the Large Magellanic Cloud (LMC) and 989 in the Small Magellanic Cloud (SMC).

This paper is organized as follows. Section~\ref{sec:Be_stars_activity} provides an overview of what is known about the variability of Be stars, from both a theoretical and observational standpoint.
The data utilized in this work are presented in Sect.~\ref{sec:sample} and in Sect.~\ref{sec:data_analysis} we introduce the methodology employed in the data analysis.
A suite of models run to aid data interpretation is presented in Sect.~\ref{sec:models}.
Our results can be found in Sect. \ref{sec:results}, followed by discussion and conclusions.

\section{Be star activity}
\label{sec:Be_stars_activity}


The number of surveys dedicated to monitoring large areas of the sky has increased significantly in the last few decades.
The primary science goals of such surveys may differ in scope (e.g. exoplanet detection, transient discoveries, microlensing searches), but the datasets that are made available to the community greatly benefit many areas of astrophysics beyond the main survey goals. 
The field of Be stars is no exception, as these surveys have provided light curves for hundreds to thousands of stars. However, with no models at the time capable of reproducing the complex dynamics of Be stars, a gap between data availability and our capacity to interpret it began to emerge. This gap is evident in the initial efforts to classify photometric variability, as seen in the works of \citet{keller_2002AJ....124.2039K} and \citet{Mennickent2002A&A...393..887M}. Both proposed similar, albeit arbitrary, categorizations of light curve features.
Since these pioneering works, the gap between the data and our understanding of the mechanisms driving variability has decreased significantly. In this section, we summarize the most important developments in our understanding of Be star activity, with an emphasis on photometric studies.

The {aforementioned gap} began to {narrow} more than a decade ago with the first attempt to use VDD theory to understand the dynamics of Be stars by \cite{Jones_2008MNRAS.386.1922J}.
Four years later, \cite{haubois_2012} presented the first systematic study of the dynamics of the Be star disk. This was made possible by combining \textsc{singlebe} hydrodynamic simulations \citep{Okazaki_2007ASPC..361..230O} with \textsc{hdust} 3D radiative transfer calculations
\citep{Carciofi_2006ApJ...639.1081C,Carciofi_2008ApJ...684.1374C} to study how different scenarios of stellar mass loss, as well as varying disk viscosity values, affect the temporal evolution of the disk and, consequently, its continuum emission.
In the context of this work, the study led to several relevant conclusions, summarized as follows:

\begin{itemize}

\item[i)] The VDD model has been conclusively shown to reproduce the photometric variations ({both in brightness} and color) of the Be star. This demonstrates that the model effectively describes the dynamics of the disk material as it diffuses inward and outward under the influence of viscous forces and in response to changes in the star's mass injection rates.

\item[ii)] It has been established that {shell} stars are equivalent to Be stars observed edge-on. For Be stars viewed at lower inclinations, the formation of a new disk appears as a net brightening and reddening of the system (hereafter referred to as a bump). In contrast, for {shell} stars, the same disk formation leads to a dimming of the system without a strong color change, hereafter referred to as a dip.
{The important effects of the inclination angle are}
discussed in greater detail in Sect.~\ref{sec:models}, as they play an important role in our analysis.

\item[iii)] 
The disk evolves in an inside-out manner, both during its growth and dissipation phases. 

\item[iv)] Different observables originate from distinct regions of the disk and therefore trace different dynamics.

\end{itemize}

This last feature is justified by two key points.
First, the viscous timescale is given by \citep{Pringle_1981ARA&A..19..137P}
\begin{equation}\label{eq:viscoustimescale}
    \tau_{vis} = \varpi^2 /\nu,
\end{equation} 
where $\tau_{vis}$ is the typical time it takes for a density perturbation to redistribute through the disk, $\varpi$ is the characteristic physical size of the fluid and $\nu$ is fluid's viscosity. It follows that the timescale is sensitive to both the disk's radial extent and its viscous properties, making it a critical parameter for understanding disk dynamics. Second, \citet{Vieira_2015MNRAS.454.2107V} demonstrated that the disk acts as a pseudophotosphere, with its apparent size increasing with wavelength. This implies that emission at shorter wavelengths originates from a compact region near the star, while emission at longer wavelengths comes from progressively larger areas of the disk.
Taken together, these two points suggest that visible photometric data should respond rapidly to variations in the mass injection rate. The shorter wavelengths probe regions closer to the star, where density perturbations are redistributed on relatively short viscous timescales, allowing for a quick observational response to changes in disk structure. This is a key point that will be further explored in the following.

A related property of viscous disk dynamics is the recently described \textit{mass reservoir effect}
\citep[MRE,][]{Rimulo_2018MNRAS.476.3555R,Ghoreyshi_2018MNRAS.479.2214G}, whereby mass accumulated in the outer parts of the disk feeds the inner disk during dissipation, substantially decreasing the rate of photometric variations. 
As noted earlier, during disk formation, material is injected near the star, where it accumulates and spreads outward through viscous redistribution. Conversely, in the dissipation phase, the depletion of material begins closest to the star and progresses outward due to the shorter viscous timescales in the inner disk. However, as dissipation advances, the mass stored in the outer disk diffuses inward, replenishing the inner regions and slowing the observational response to the dissipation process.
This interplay between the inner and outer disk is a key characteristic of viscous disks and will be central to the discussion in the following sections.


Long-term photometric studies also have made significant contributions and probably will {play a major role} in a more complete understanding of the Be phenomenon. Two recent examples are 
\cite{Labadie-Bartz_2017AJ....153..252L} and \cite{Bernhard_2018MNRAS.479.2909B}.
In the methodology developed by \citeauthor{Labadie-Bartz_2017AJ....153..252L} (and followed by \citeauthor{Bernhard_2018MNRAS.479.2909B}), light curves are inspected for the presence of outbursts, i.e., a clear and sudden departure from the photometric baseline followed by a gradual return to it \citep{Sigut_2013ApJ...765...41S}.
Variations were grouped into three ranges: short ($<2$ days), intermediate ($2$ to $200$ days) and long term ($>200$ days).
Variations on long and intermediate timescales are typically linked to changes in the disk structure \citep[as summarized above; see][for further details]{haubois_2012}, while short-timescale variations are generally dominated by non-radial pulsations \citep[see, e.g.,][]{Labadie-Bartz2022AJ_163_226}.

Analyzing a sample of 610 light curves of Galactic Be stars observed by KELT (Kilodegree Extremely Little Telescope, \citealt{Pepper_2007PASP..119..923P}), \citet{Labadie-Bartz_2017AJ....153..252L} found that: clear outbursts occurred in 20\% of the stars; features consistent with non-radial pulsations were identified in 25\%; 37\% exhibited long- or intermediate-timescale variations. Additionally, their study revealed that early-type stars are generally more active, displaying a higher frequency of outbursts.
Similar findings were reported by \citet{Bernhard_2018MNRAS.479.2909B}, who additionally identified long-term variability in $36 \pm 6\%$ of their sample. Outbursts were observed in $73 \pm 5\%$ of the light curves analyzed from ASAS-3 data (All-Sky Automated Survey-3; \citealt{Pojmanski_2002AcA....52..397P}), which included 287 Galactic Be stars (233 confirmed and 54 candidates).
{One may note that longer observational (and lower precision) baseline surveys are well-suited to capture long-lived events.
In contrast, shorter duration, at higher-precision surveys excel at detecting faster, lower-amplitude events but may miss longer-duration phenomena.}

\citet{Jian_2024A&A...682A..59J} also made relevant contributions to the study of Be star activity, discovering 736 new Be stars.
Based on WISE data (Wide-field Infrared Survey Explorer), the authors studied the typical behavior in a mid-infrared color-magnitude diagram (CMD) of a sample of known Be stars, from which it was possible to establish color-brightness variation criteria used to identify new candidates.
There are two main advantages to studying Be stars in the infrared: the amplitude of 
{bumps and dips are} 
expected to be larger in the IR than in the visible, and the interstellar extinction is weaker as well. The Be nature of the candidates was confirmed by LAMOST (Large Sky Area Multi-Object Fiber Spectroscopic Telescope) $H_\alpha$ observations.
The sample studied was composed of light curves for 916 stars, presenting a typical duration of disk build-up of $474$ days and a decay that lasts $524$ days.
{Given the WISE cadence, it favors observations of relatively longer events.}

\section{The Sample}
\label{sec:sample}



\begin{figure*}[!t]
    \centering
    \includegraphics[width=.8\paperwidth]{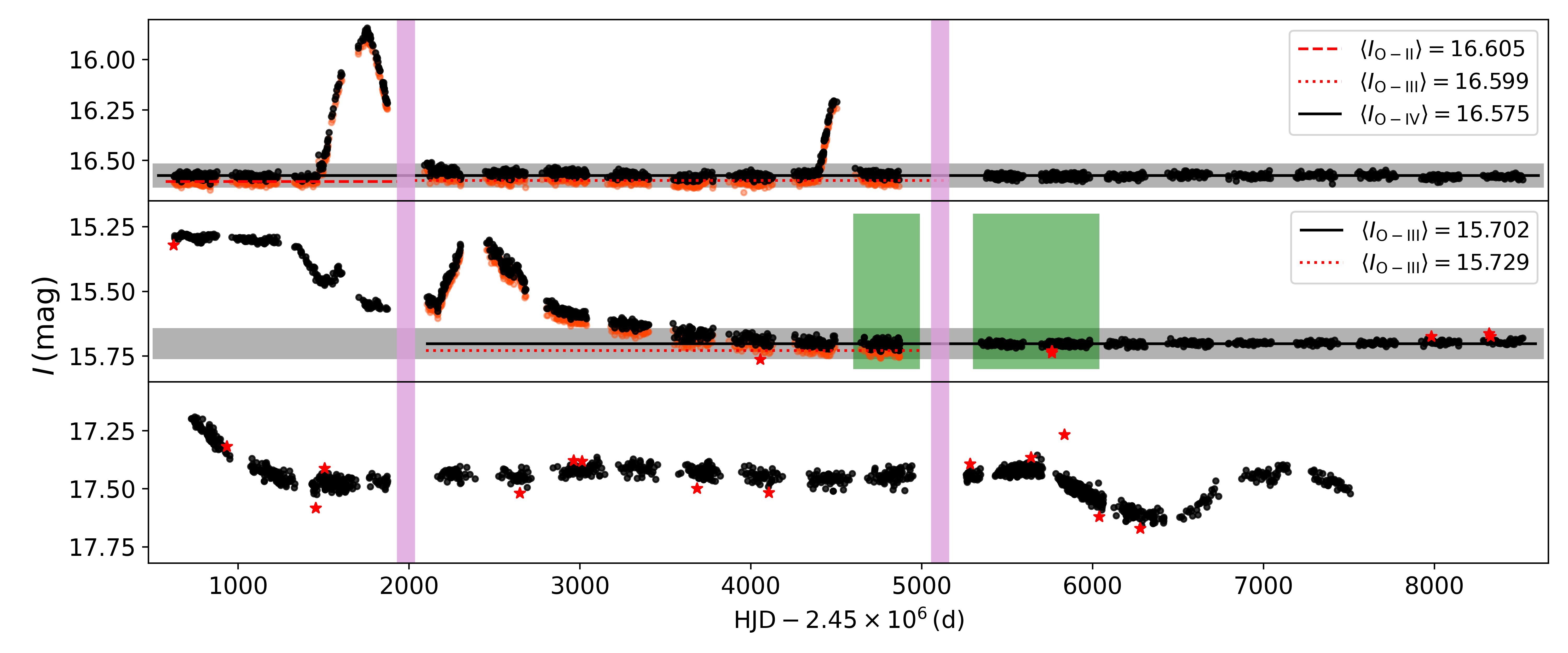}
    \caption{Examples of pre-processing steps applied to the OGLE data.
    The original data for OGLE-II and OGLE-III are plotted in red, while black points represent the adjusted light curve. The median valued of the sections with low disk activity (top) and selected by user (middle) are shown in the legends and are graphically represented by horizontal lines (red ones for original OGLE-II and OGLE-III data, while black represents OGLE-IV level).
    The identified outliers are plotted as red stars.
    and gray horizontal bands display the range of two times the detection threshold of $0.06\,{\rm mag}$ (see Sect.~\ref{sec:raw_sample} for more details). 
    Vertical purple bands mark transition times among different observational phases.
    \textit{Top:} SMC\_SC3\_71445 light curve, with automatic offset values of $\Delta_\mathrm{O-II} =16.575 - 16.605 = -0.030$ and $\Delta_\mathrm{O-III} =-0.024\,\mathrm{mag}$.
    \textit{Middle:} data for SMC\_SC10\_94551 target, with offset valued equal to $\mathrm{\Delta_{\rm O-II}} = -0.027\,{\rm mag}$, estimated from manual selection (outlined by green areas), and outliers identified by sigma clipping with the parameters $\Sigma = 2\sigma$ and $N_{\mathrm{I}}=3$.
    \textit{Bottom:} LMC\_SC15\_294 light curve with outliers identified by sigma clipping ($\Sigma = 3\sigma$ and $N_{\mathrm{I}}=3$).}
    \label{fig:example_offset}
\end{figure*}

\begin{table*}[!t]
\centering
\setlength\tabcolsep{5.5pt}
\caption{Ten first objects from LMC and SMC samples, where lines are ordered by OGLE-II ID. In order, the columns represents coordinates, OGLE-II, OGLE-III and OGLE-IV IDs.
The offset type is shown by the seventh column (where A is for automatic and M means manual, as in the examples of Fig.~\ref{fig:example_offset}).
The number of removed outliers are shown in the last column.
This table is available in its entirety in a machine-readable form in the online journal. A portion is shown here for clarity.
}
\label{tab:sample_master}
\begin{tabular}{cclllcc}
\hline
\hline
RA & DEC & OGLE-II ID & OGLE-III ID & OGLE-IV ID & Offset & Rem. Outliers \\
$(^\circ)$ &  $(^\circ)$ & & & & & \\
\hline
9.86404 & -73.55267 & SMC\_SC1\_101159 & SMC125.8.20775 & SMC713.02.15158 & - & 2 \\ 
9.74837 & -73.42361 & SMC\_SC1\_105769 & SMC125.7.4292 & SMC713.10.30 & M & 8 \\ 
9.84204 & -73.42544 & SMC\_SC1\_105816 & SMC125.7.4358 & SMC713.10.252 & A & 9 \\ 
9.77983 & -73.34981 & SMC\_SC1\_108209 & SMC125.7.24762 & SMC713.10.54 & - & 13 \\ 
9.18796 & -73.55000 & SMC\_SC1\_11026 & SMC130.1.9088 & SMC713.03.13457 & A & 10 \\ 
9.21912 & -73.48242 & SMC\_SC1\_12973 & SMC130.1.9028 & SMC713.03.13438 & M & 18 \\ 
9.08971 & -73.43628 & SMC\_SC1\_12977 & SMC130.2.9346 & SMC713.11.5 & A & 4 \\ 
9.08096 & -73.43139 & SMC\_SC1\_15170 & SMC130.2.10411 & SMC713.11.1175 & A & 1 \\ 
9.31442 & -73.76264 & SMC\_SC1\_32299 & SMC133.4.8792 & SMC720.16.11391 & A & 13 \\ 
9.09596 & -73.78083 & SMC\_SC1\_4659 & SMC133.3.10134 & SMC720.16.13955 & A & 6 \\ 
\hline
\hline
83.31325 & -70.36894 & LMC\_SC1\_108086 & LMC170.3.21 & LMC516.04.19670 & - & 53 \\ 
83.44342 & -70.33486 & LMC\_SC1\_113297 & LMC170.3.73425 & LMC516.04.19664 & A & 13 \\ 
83.40288 & -70.16544 & LMC\_SC1\_124623 & LMC170.4.81503 & LMC516.12.2 & A & 19 \\ 
83.32333 & -70.20175 & LMC\_SC1\_124756 & LMC170.4.64358 & LMC516.12.274 & A & 2 \\ 
83.31079 & -70.18892 & LMC\_SC1\_124784 & LMC170.4.64070 & LMC516.12.324 & - & 12 \\ 
83.30254 & -70.16481 & LMC\_SC1\_124845 & LMC170.4.64064 & LMC516.12.413 & - & 15 \\ 
83.33104 & -70.09783 & LMC\_SC1\_136951 & LMC169.1.16837 & LMC516.12.12288 & - & 26 \\ 
83.33754 & -70.10039 & LMC\_SC1\_136999 & LMC169.1.16867 & LMC516.12.12344 & A & 14 \\ 
83.25717 & -70.37517 & LMC\_SC1\_14210 & LMC170.3.15 & LMC516.04.19669 & - & 21 \\ 
83.16358 & -70.3545 & LMC\_SC1\_14373 & LMC170.6.112550 & LMC516.04.64676 & A & 5 \\ 
\hline
\end{tabular}
\end{table*}

In this section, we present a brief overview of the analyzed data. Since it consists of multiple observational phases from a ground-based photometric survey, we also detail the procedures used to create a homogeneous sample for examination.

Our sample is composed of OGLE (Optical Gravitational Lensing Experiment) data in three photometric bands: $B$, $V$ and $I$, listed in increasing order of coverage density.
%
This endeavor, still on going, was initially conceived to search for dark matter in the Galaxy using microlensing \citep{Paczynski_1986ApJ...304....1P}, making it part of the first generation of microlensing projects, alongside MACHO \citep[Massive Compact Halo Object, ][]{Alcock1997ApJ...486..697A} and EROS \citep[Exp\'erience pour la Recherche d'Objets Sombres,][]{Aubourg1993Natur.365..623A}.
During its initial phase, a 1-meter Swope telescope was used. This was later upgraded to a dedicated 1.3-meter telescope, the Warsaw Telescope, constructed at Las Campanas Observatory, which achieved first light in 1996.
Its first observational phase, which lasted from 1992 to 1995 \citep{Udalski_1992AcA....42..253U}, was followed by three more observational phases, namely OGLE-II \citep[from 1997 to 2000,][]{Udalski_1997AcA....47..319U}, OGLE-III \citep[2001 to 2009,][]{udalski2008AcA_58_329U}, and OGLE-IV \citep[2009 to 2025,][]{udalski_2015AcA....65....1U} and an ongoing 5th phase.
Over the past 30 years, both the instrument and the data reduction process have been significantly improved, enhancing the quality of the delivered data and achieving photometric accuracy on the order of milli-magnitudes in its latest phase. These advancements have also allowed the project's goals to expand, encompassing the detection and characterization of variable stars, dwarf novae, and studies of the structure of the Galaxy and the Magellanic Clouds.
%

The Be star candidates were selected by \citet{Mennickent2002A&A...393..887M} and \citet{Sabogal2005MNRAS.361.1055S} for the LMC and SMC, respectively. 
Both selections rely on color and magnitude criteria derived from OGLE-II data. Spurious variables such as binaries and Cepheids were removed by visual inspection.
For this project, the third and fourth observational phases (OGLE-III and OGLE-IV) were incorporated into the above sample, resulting in light curves {covering 23 years}: 1997 to 2000 (OGLE-II); 2001 to 2009 (OGLE-III) and 2010 to 2020 (OGLE-IV) \citep{udalski_2015AcA....65....1U}.
Given that our sample consists of observations that span more than two decades and was collected using three different setups, several steps were taken to ensure that the sample was as homogeneous as possible. These procedures are detailed below.

\subsection{Cross-matching between OGLE phases}

{This step is necessary because OGLE sources, and their brightness and coordinates, are internally defined based on the observed PSFs on the CCD, in contrast with more recent surveys, where lightcurves are extracted using coordinates based on other catalogs.
}
To combine all three observational phases, a cross-match was performed using the OGLE-II coordinates as a reference. An error margin of $3$ arcseconds was adopted, ensuring that the distance between sources in multiple phases did not exceed this limit.
The $I$-band data, which have the highest observational cadence, was used as a reference.
To minimize false matches in dense fields, the median brightness level in this band was used as a baseline, with brightness differences in subsequent phases required to stay within a limit of $1.0$ mag.
Correspondence was found for all targets; however, the following cases required special attention:

\begin{enumerate}
    \item Unique match: for most of the sample (2840 stars, 90.6\%) a unique match was found in all observing phases.
    
    \item Partial match: For a small number of targets (44 stars, 1.26\%), no corresponding data was found in the OGLE-IV dataset, resulting in a smaller observing baseline (about 12 years)
    
    \item Multiple matches: For 7.9\% (249 stars) more than one match was found for different observing phases. Within a given phase, multiple matches showed the same variability, suggesting the existence of stars with duplicate (or multiple) identifiers in OGLE. These were removed by visual inspection.
\end{enumerate}

\subsection{Correction of photometric offsets}

For some stars, shifts in the photometric level are observed between different OGLE phases, which are particularly notable in light curves with lower activity. These shifts can introduce systematic errors into the combined light curves.
To address this, we construct a seamless light curve by applying offsets to the different phases. The procedure involves using the photometric calibration of OGLE-IV, which is more accurate \citep{udalski_2015AcA....65....1U}, as a reference to align the other two phases.
{
We employed two different ways to estimate the offset to be applied, depending on the characteristics of the light curve. 
{The goal is to identify the baseline magnitude level in each phase. This is done by finding the data in the 3\textsuperscript{rd} percentile (or 97\textsuperscript{th} for edge-on), and taking the median. These median values for the different phases are then offset to match so that the entire LC has a consistent baseline level.}
%
Alternatively, if the baseline is present in two observational phases, the same calculation is performed using time windows selected by the user.}

The top two panels of Fig.~\ref{fig:example_offset} show examples of both procedures. The original data for star SMC\_SC3\_71445 
are shown in red in the top panel for OGLE-II and OGLE-III, while the final light curve is plotted in black for all phases.
The 3\textsuperscript{rd} percentile for all phases is used to automatically offset phases, where red lines show the original median values of the first two observational phases, black line represents the median value of phase OGLE-IV, as indicated at the legend box.
The second panel shows an offset example for target SMC\_SC10\_94551, where a baseline phase is missing in OGLE-II and the above procedure was repeated using the selected time window, outlined by green areas that correspond to $t = \mathrm{HJD} - 2.45 \times 10^6 = 4700$ to $4950$ and $5200$ to $6050\,{\rm d}$, 
approximately.

\subsection{Removal of outliers }

Many light curves possess clear outliers, i.e., 
data points diverging from the neighboring points by values much larger than the local variance of the data, frequently presenting differences of one magnitude or more.
To remove outliers, we choose to use a sigma-clipping function from the \textsc{astropy}\footnote{\url{https://www.astropy.org}} package that allows for different parameters such as the number of iterations ($N_{\rm I}$) and error level ($\Sigma$).
The bottom part of Fig.~\ref{fig:example_offset} shows an example of this process for the star LMC\_SC15\_294.
The sigma-clipping function was run with $N_{\rm I} = 2$ and $\Sigma=3\sigma$ resulting in 13 identified outliers, represented by red stars.

\subsection{Raw sample}\label{sec:raw_sample}


\begin{figure*}
    \centering
    \includegraphics[width=.8\paperwidth]{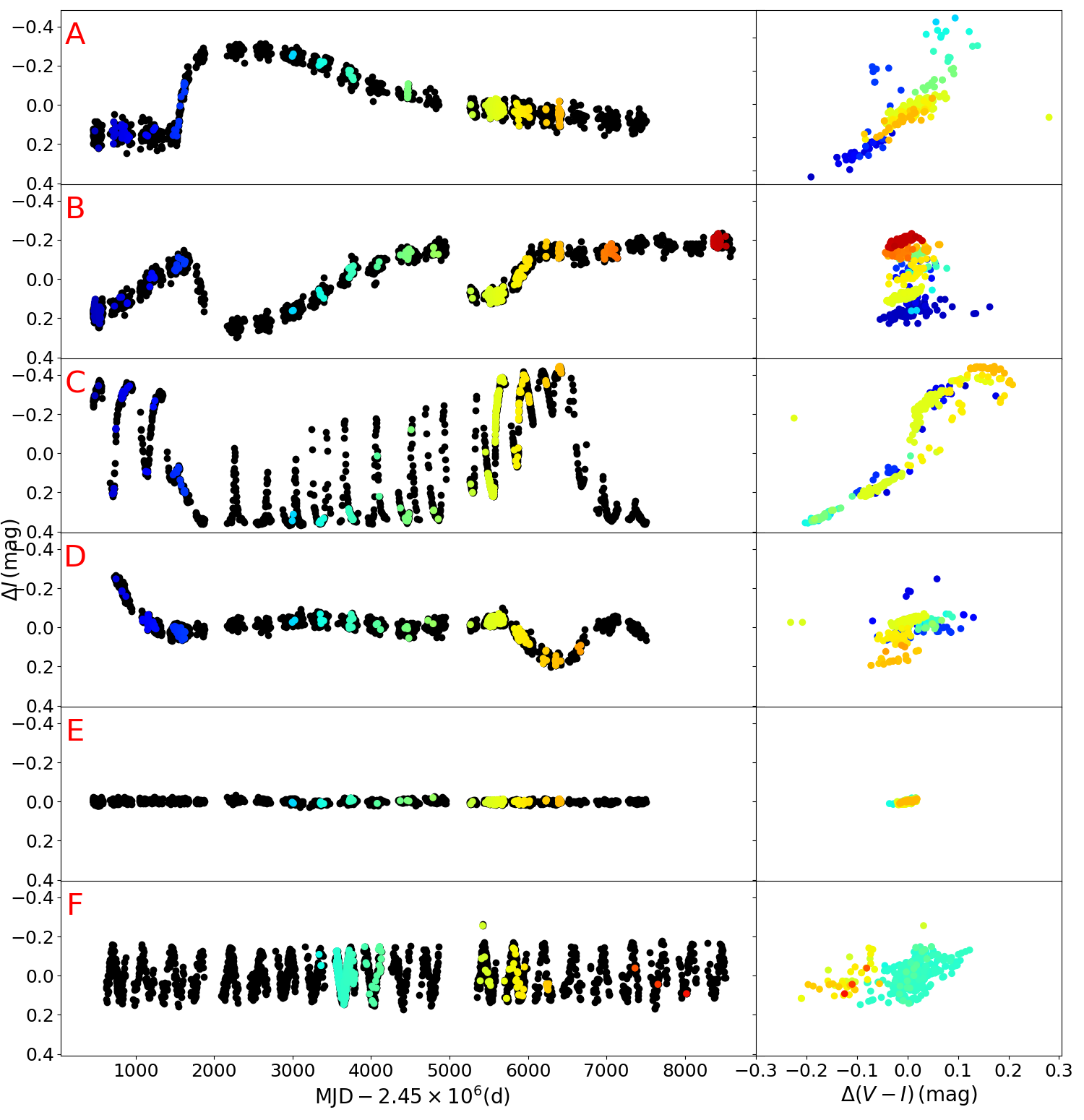}
    \caption{Light curve sample archetypes.
    In the first panel (A) it is shown the LMC\_SC1\_164770 light curve with initial stable phase followed by a bump with a five thousand days long decay, with a excursion in the CMD draws a loop, leaving a bluer and dimmer region, going through brighter and redder regions and falling back to initial region.
    The second light curve, from target LMC\_SC3\_400956, begins with a slow brightness increase and is followed by two dips; while the associated CMD shows a vertical excursion with small changes in color.
    Panel C (LMC\_SC1\_340178) presents a very active star with (quasi)-periodic bumps and a CMD structure similar to one find in the first panel.
    In the next panel displays the light curve of LMC\_SC15\_294, exhibiting a more complex behavior is presented; the CMD do not present a clear structure.
    The light curve presented in panel E (LMC\_SC3\_113787) shows no activity and small CMD variations.
    The last panel presents SMC\_SC2\_33373 a quite periodic star with no clear CMD structure.
    }
    \label{fig:examples_LCs}
\end{figure*}

After applying the described procedures, we obtained a raw sample of 3,133 light curves spanning nearly 20 years, with 989 objects from the SMC and 2,144 from the LMC.
Offsets and outlier removal were performed for 79.3\% (1,700) and 10.9\% (234) 
of LMC targets, respectively, and for 52.8\% (522) and 5.7\% (56) of SMC light curves.
Targets missing OGLE-IV data account for 18 (0.84\%) in the LMC and 78 (7.9\%) in the SMC. 
Combined, the entire sample encompasses over sixty millennia of photometric data.

A key concept for this work is what we defined as \emph{baseline}.
For purposes that will be clear in the following sections, a star's baseline is the state in which there is no circumstellar matter \citep[this state was referred to as inactive by][]{2016A&A...593A.106R}.
Therefore, the brightness and color measured for a Be star in this phase are reddened photospheric values.
To go from the later value to intrinsic brightness and colors, we need interstellar reddening information that was obtained from the reddening maps provided by 
\citet{Skowron2021ApJS..252...23S}.
%
Based on Red Clump color properties from the OGLE-IV data, the authors apply different methods to obtain the reddening value (their table 1).
We then cross-matched the reddening maps with targets from our sample that display baseline features. 
The resulting intrinsic $I$ and $V-I$ values will be compared with photospheric models from both metallicities (Sect.~\ref{sec:model_disk-less})
in Sect.~\ref{sec:results_Mass}, below.
%
Additionally, baseline phases are used to estimate the detection threshold of OGLE. Details are given in Appendix~\ref{sec:app_observationalthreshold}.
The value obtained was $0.03\,{\rm mag}$, but we adopted twice this limit for classification purposes. The usage of the detection threshold is exemplified in the next section.

As expected from the erratic nature of Be stars, our sample contains a wide variety of behaviors.
In Fig.~\ref{fig:examples_LCs}, we show a visual summary of patterns found in $I$-band light curves and their respective $V-I$ CMDs.
This figure highlights well-behaved stars (e.g., panels A, B, and E), highly active stars (panels C, D, and F), and stars exhibiting (quasi-)regular {variability} (panels C and F).
The $V-I$ CMDs also show diverse behaviors, with some stars following reddening tracks while others displaying little color variations.
The literature offers two approaches to samples like this: a quantitative one, such in the works of \citet{Labadie-Bartz_2017AJ....153..252L}, \citet{Bernhard_2018MNRAS.479.2909B}, and \citet{Jian_2024A&A...682A..59J} or a modeling approach, where each light curve is reproduced in detail, as exemplified by
\citet{Rimulo_2018MNRAS.476.3555R} and \citet{Ghoreyshi_2018MNRAS.479.2214G}.
We pursue a third approach here, combining elements of both alternatives.
Specifically, we propose a dynamic classification that identifies phases where mass loss is occurring, periods when it is zero or negligible, and epochs indicating the presence or absence of a disk, among other features.
A dynamic classification provides deeper insights into the data compared to the quantitative approach and is more practical than the modeling alternative, given the size of our dataset.
However, this approach relies on the use of VDD theory as a framework for interpreting light curves. A significant challenge is estimating the inclination angle of 
{the Be stars in our sample based on the photometric brightness and color variations alone}, which is crucial for identifying the dynamic phases of a light curve.
In the next section, we detail how a set of models aids in determining stellar inclinations, providing insights into the example light curves shown in Fig.~\ref{fig:examples_LCs}. This forms the basis for an alternative analysis, which is presented in Sect.~\ref{sec:data_analysis}.

\section{Models as an interpretative  key}\label{sec:models}

This work employs two families of models as interpretative tools to deepen our understanding of the data sample and extract maximum information.
In Sect.~\ref{sec:models_dynamic}, we introduce dynamical models that aid in distinguishing the inclinations of light curves using color information.
The second set of models, discussed in Sect.\ref{sec:model_disk-less}, focuses on diskless Be stars and allows us to estimate stellar parameters, particularly mass, for light curves meeting specific criteria.

\subsection{Dynamical Models}
\label{sec:models_dynamic}

Two primary challenges in studying the light curves of our sample are the limited information regarding the fundamental parameters of the central star (e.g., mass, rotation rate) and the uncertainty about its inclination angle. While the former is addressed in the following section, the latter is discussed below.

A possible way to discriminate orientation is by using color information, an observational fact first identified by \citet{Harmanec_1983HvaOB...7...55H} and later explored by \citet{haubois_2012}.
Inspired by the latter work, we use the same tools that they used to run a set of dynamical simulations, tailored for the LMC Be stars, to explore how the observational signature changes as a function of the inclination angle in response to disk events.
It is important to mention that our results are quantitatively different from those of \citet{haubois_2012}, since they studied the photometric signature in the visible ($V$ band), while we focus on the brightness variations in the $I$ band, which comprises the bulk of the OGLE data.

The models were computed using the \textsc{singlebe} \citep{Okazaki_2007ASPC..361..230O} code coupled with the NLTE radiative transfer code \textsc{hdust} \citep{Carciofi_2006ApJ...639.1081C}.
\textsc{singlebe}'s main task is to solve the time-dependent evolution of an isothermal VDD, using a thin disk approximation \citep{bjorkman_e_carciofi_2005}, while \textsc{hdust} is used to solve the radiative transfer, mimicking the SED's time evolution of a star-disk system.
Since we aim to compare these models with OGLE light curves, the emergent SED was convolved with $B$, $V$, and $I$ photometric band-passes.

\begin{table}
    \centering
    \begin{tabular}{rcccc}
        Parameter & Values & & & \\\hline
        Mass $(M_{\odot})$ & 12 & & &  \\
        $R_{\rm eq.}\,(R_{\odot})$ & $6.35$ & & & \\
        $R_{\rm pole}\,(R_{\odot})$ & $4.78$ & & & \\
        $L\,(L_{\odot})$ & $13267$ & & &\\
        $W$              & $0.81$ & & & \\
        $\alpha$               & $1.0$ & & & \\
        $t_{\rm b}\,({\rm d})$                & $60$ & $180$ & $600$ & $1800$ \\
        $\Sigma_0\,({\rm g\, cm^{-2}})$ & $0.12$ & $0.69$ & $1.67$ & $4.00$\\
    \end{tabular}
    \caption{Parameters used for the dynamical models.
    }
    \label{tab:dynamicmodelsparameters}
\end{table}

The main simulation parameters used are summarized in Table~\ref{tab:dynamicmodelsparameters}.
To represent a typical Be star, we fixed the central star mass at $12\,M_{\odot}$,  with a rotation rate of $W = 0.81$, equatorial and polar radii of $R_{\rm eq} = 6.35\,R_\odot$, $R_{pole} = 4.78\,R_\odot$, respectively, while the external radius 
 is $40\,R_{\rm eq}$ and LMC metallicity.
{These parameters were chosen based on the best available data for Be stars. The mass value aligns with the peak Be star fraction observed in clusters \citep[e.g.][]{Martayan2006A&A...452..273M,Martayan_2007A&A...462..683M,Navarete2024ApJ...970..113N}, while $W$ corresponds to typical values expected for Galactic Be stars \citep{Rivinius2013A&ARv..21...69R}.}
We chose a simple dynamical scenario in which an inactive star undergoes mass loss for the duration of $t_{\rm b}$, exploring the values presented in Table~\ref{tab:dynamicmodelsparameters}.
Disk formation is followed by a dissipation phase, during which there is no mass loss, lasting for $t_{\rm d} = 4\times t_{\rm b}$.
In addition,  
we varied the disk base density, $\Sigma_0 $, assuming values ranging from $0.12$ to $4.00\,\mathrm{g}\,\mathrm{cm}^{-2}$. 
$\Sigma_0$ is the density at $r=R_{\rm eq}$, achieved following an infinitely extended phase of mass loss.
The values adopted for those parameters cover the range found by \citet{Vieira_2017MNRAS.464.3071V} for Galactic Be stars.
{All models were run using the same value for the viscosity parameter} $\alpha = 1.0$ and for thirty observers with viewing angles evenly distributed in $\mu = \cos(i)$ between 0 and 1.

\begin{figure}
    \centering
    \includegraphics[width=0.49\textwidth]{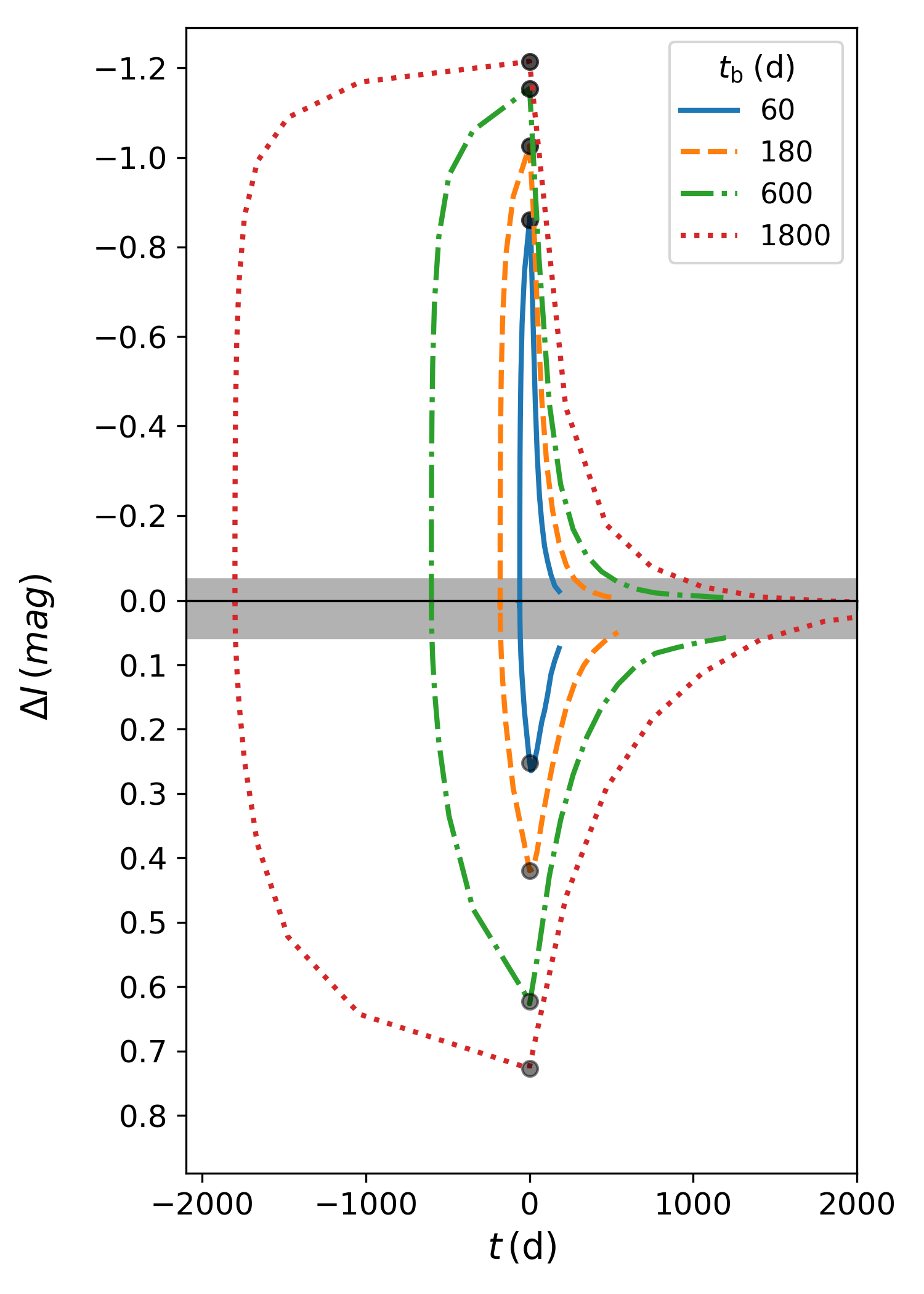}
    \caption{
    Synthetic light curve of disk events for a dense disk ($ \Sigma_0 = 4.00\, {\rm g\, cm^{-2}}$), generated for different build-up times as indicated in the legend.
    The abscissa represents time, referenced to the beginning of dissipation, while the ordinate shows the variation in brightness, relative to the diskless value, caused by the disk's presence.
    The gray band represents the detection threshold, as introduced in Fig.~\ref{fig:example_offset}.
    The transition to dissipation phase is marked by black  circles.
    The upper panel shows a pole-on view (${i = 0^\circ}$), while the edge-on view (${i = 90^\circ}$) is in the bottom.
    }
    \label{fig:modelos_bumps_tempos}
\end{figure}

Figure~\ref{fig:modelos_bumps_tempos} shows how the photometric signature in the $I$ band changes according to the build-up times, with the upper panel showing the lightcurve seen by a pole-on observer and the bottom panel representing the equivalent for an edge-on viewer.
Notably, the lightcurves do not reach saturation, even in models with the longest feeding time, $t_{\rm b} = 1800\,{\rm d}$, and pole-on viewing,
suggesting that a steady state is not reached in the parts of the disk whence most of the $I$-band excess flux comes from.
The edge-on models are even farther from saturation. Unlike the pole-on case, where the $I$-band variations can be interpreted as originating from a pseudophotosphere with a varying radius \citep{Vieira_2015MNRAS.454.2107V},  the edge-on orientation is influenced by absorption throughout the entire disk. Considering the viscous timescale (Eq.~\ref{eq:viscoustimescale}), it is evident that the outer regions of the disk take longer to reach a steady state.

During dissipation, the balance between emission and absorption plays a critical role.
For example, excess emission associated with disk presence falls below the detection threshold after $\sim 750$ days for the longer simulation. However, the corresponding model still shows signs of dissipation when seen edge-on. The same trend is also observed for the other models. 
It is remarkable that the inclination angle, $i$, plays such a fundamental role in the light curves, controlling not only the amplitude {and sign} of the photometric signature but also its rate of variation.
Another key aspect of Fig.~\ref{fig:modelos_bumps_tempos} is the previously mentioned {MRE, according to which the longer the duration of mass loss (and consequently, the greater the disk mass that has been built up), the longer the dissipation phase \citep{Rimulo_2018MNRAS.476.3555R,Ghoreyshi_2018MNRAS.479.2214G}.}

\begin{figure}
    \centering
    \begin{tabular}{c|c}
         \includegraphics[width=0.45\textwidth]{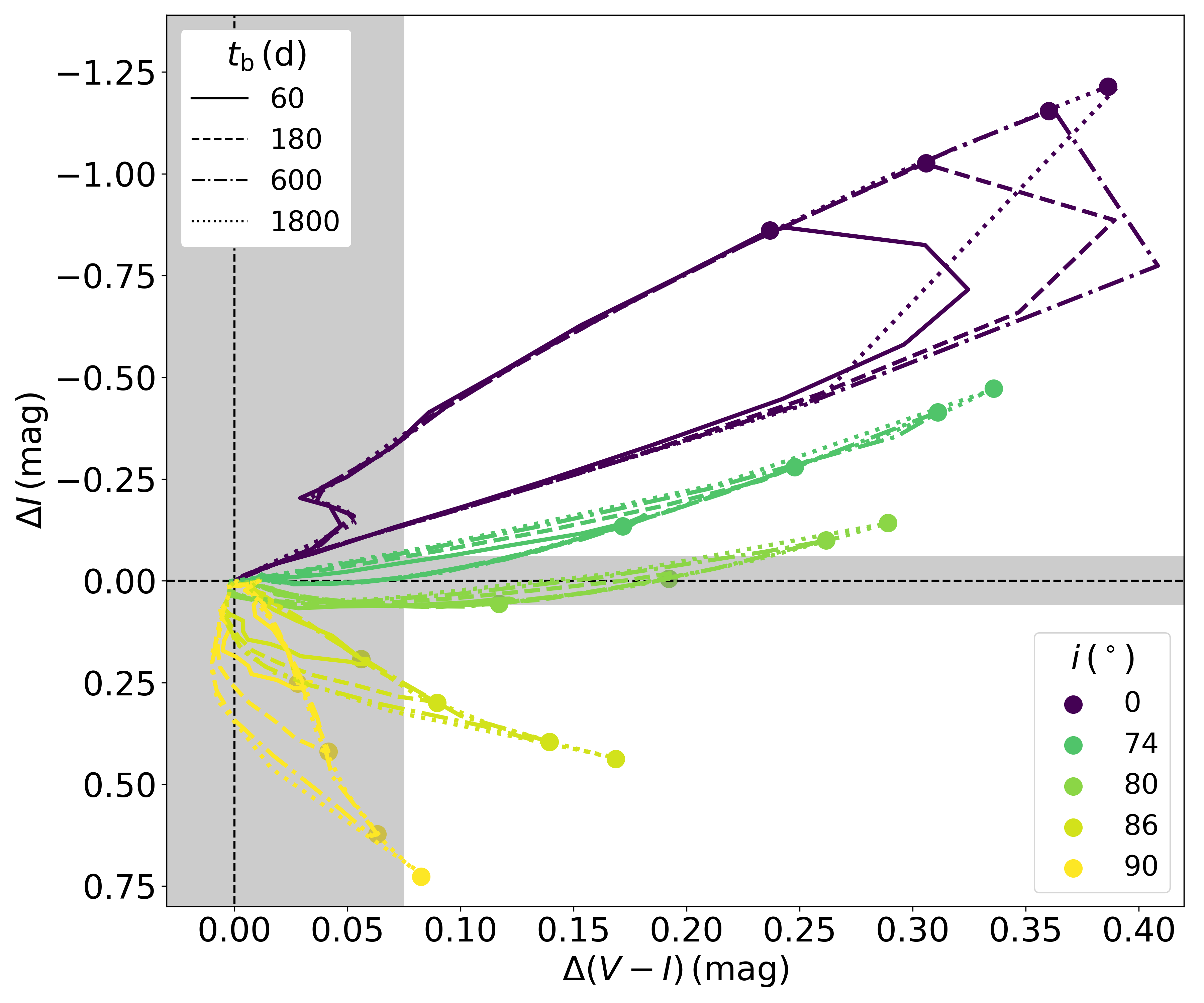} \\
         \includegraphics[width=0.45\textwidth]{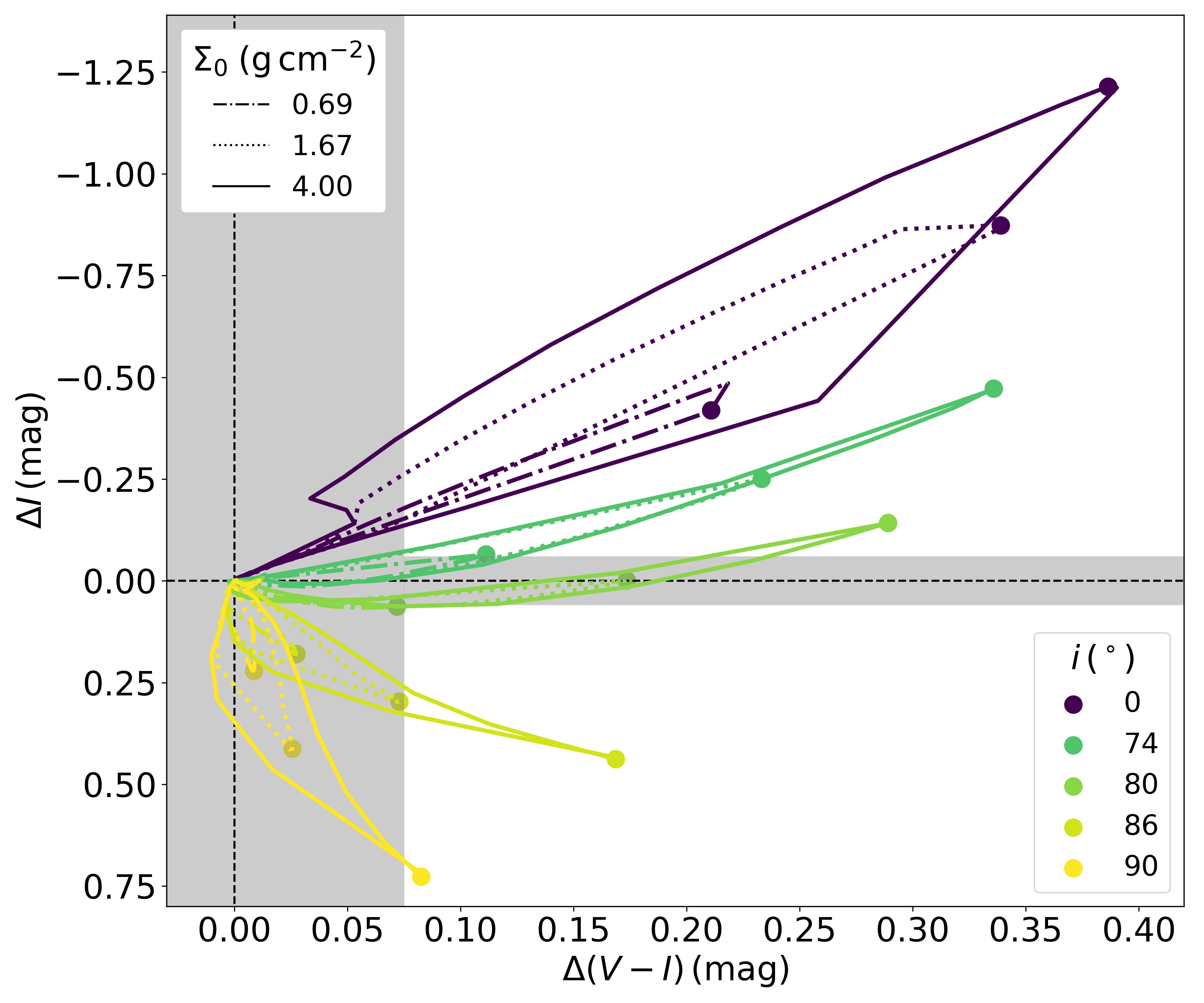}
    \end{tabular}
    \caption{
    $V-I$ CMD for different models and inclinations. The ordinate shows the brightness variations in the $I$ band, taking a diskless star as reference. The abscissa represents the change in color due to disk’s presence, with zero indicating a diskless star. 
    \textit{Top}: Shown are models for $\Sigma_0 = 4.00\,{\rm g\, cm^{-2}}$, selected build-up times (different line styles as seen in the top left legend) and  inclination angles (indicated by different colors in the bottom right legend).
    \textit{Bottom}: Same as the top panel but keeping $t_{\rm b} =1800\,{\rm d}$ fixed and varying the base density.
    The transition to dissipation phase is marked by filled circles, colored according to $i$.
    Dashed horizontal and vertical lines represent the CMD position for a diskless star.
    As in Fig.~\ref{fig:example_offset}, the horizontal gray area represents the adopted detection threshold, while vertical gray band indicates the equivalent limit for the color, taking into account the error propagation error for both band-passes.}
    \label{fig:modelos_CMD_tempos}
\end{figure}

The top part of Fig.~\ref{fig:modelos_CMD_tempos} shows the paths traced in the $V-I$ CMD for high-density models.
To facilitate visualization, we selected models with build-up times of 60, 180, 600 and 1800 d for a few values of inclination angle.
All models follow a loop in this figure, qualitatively similar to those reported by \citet{haubois_2012}.
For pole-on viewing, the build-up phase is characterized by a top-right path in the figure, indicating that the system becomes redder as the disk grows.
When dissipation starts, the loop returns to its starting point. 
As the observer's inclination angle moves from $0^\circ$ to $90^\circ$, the orientation of the loop rotates clockwise, and the transition point shifts from a brighter and redder position toward dimmer magnitudes with decreasing color variation.
The $80^\circ$ model presents an interesting feature: while displaying large color variations, it shows very little changes in brightness. Similar behavior was reported by \citet{haubois_2012}, and bears important consequences for this work, as discussed below.
For edge-on models, the variation is more significant in brightness than in color, with the latter varying by less than $0.1$.
Longer-fed disks present wider loops that reach farther from the starting point.
The behavior for less dense models are qualitatively similar. All corresponding plots are shown in Appendix~\ref{sec:app_CMDlowdensity}.

\begin{figure}
    \centering
        \includegraphics[width=0.49\textwidth]{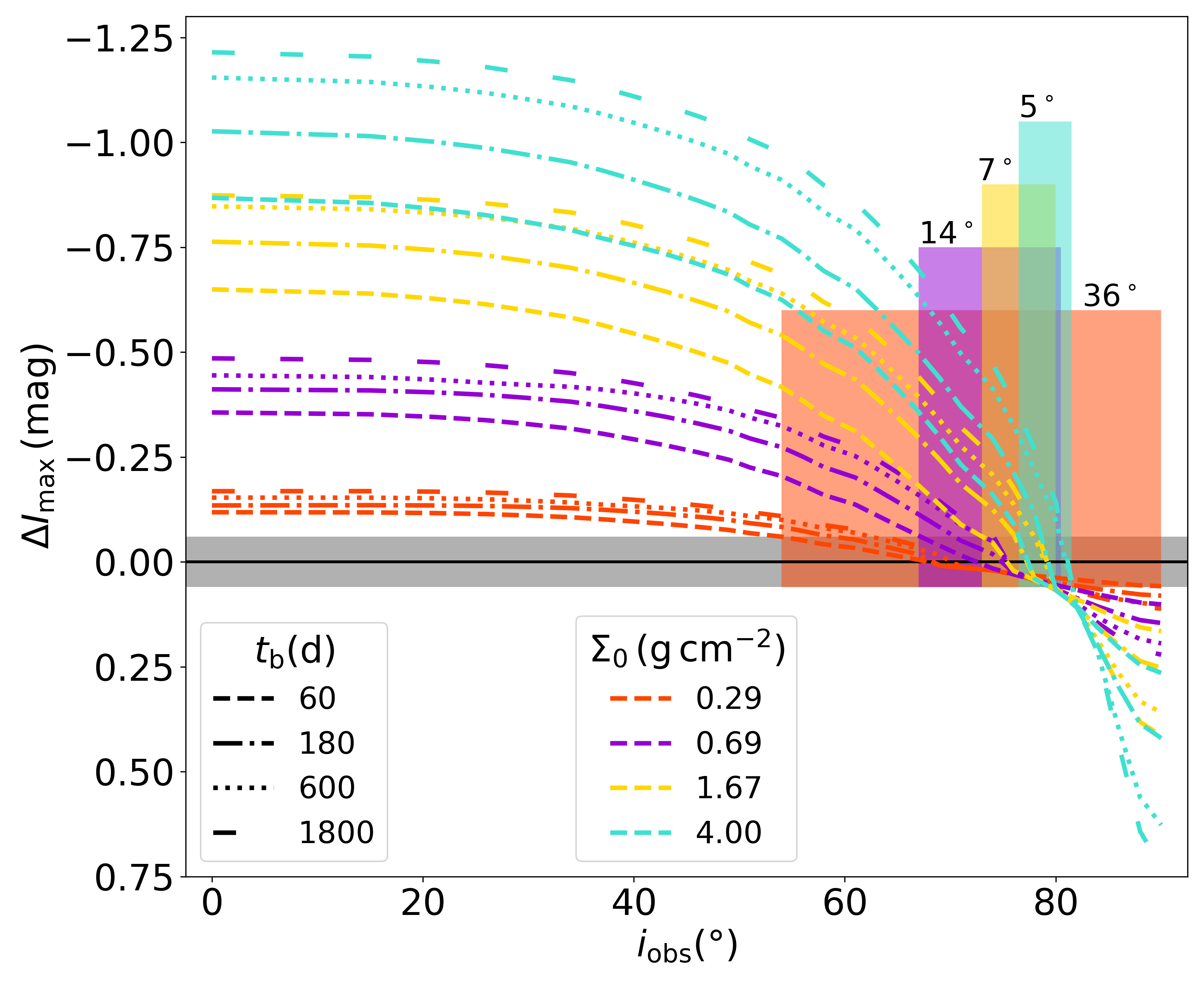}

    \caption{
    Maximum disk excess for the $I$ band, $\Delta I_{\rm max}$, observed as function of inclination angle for models with different disk densities (represented by colors shown in the central legend box) and feed times (line styles as seen in the
    bottom left legend). 
    {The horizontal gray band indicates the detection threshold limit of $0.06\,{\rm mag}$.}
    The vertical bars indicate the inclination range for which $\Delta I_{\rm max}$ is below the detection limit, colored accordingly the model density and with the respective numerical range at the top.
    }
    \label{fig:modelos_angulointermediario}
\end{figure}

Although our models assume a simplified dynamical scenario, the top panel of Fig.~\ref{fig:modelos_CMD_tempos} suggests that, except for intermediate-angle cases, the loop's orientation can serve as a diagnostic for the inclination angle.
The natural question that arises is: How well can the inclination angle be diagnosed based on the observed slope of the loop?
From Fig.~\ref{fig:modelos_CMD_tempos}, we see the important fact that the orientation of the loop does not depend on the build-up time, for a given base density (top panel). However, it does depend on the base density (bottom panel).

The limited ability to extract information about the inclination angle from lightcurves is further demonstrated in Fig.~\ref{fig:modelos_angulointermediario} that plots the maximum magnitude change of a model vs. inclination angle.
The figure enables comparisons between models with the same $\Sigma_0$ (represented by lines of different colors) and varying $t_b$ (indicated by different line styles).
All models show similar behavior: the maximum flux excess is always seen for $i=0^\circ$, reaching almost $1.25\,{\rm mag}$ in the $I$ band for the densest and longest built disk. The excess drops slowly as $i$ grows, in some cases being almost constant between 0 and approximately $50^\circ$.
The rate of variation increases substantially at this point, with the excess reaching zero and then becoming negative at inclinations between approximately $60^\circ$ to $80^\circ$, depending on how long the disk is fed and its density.
The figure also highlights
the range of inclination angles where the models produce excess below the detection threshold defined earlier. 
{The numbers above the bar show the respective inclination range encompassed by each bar.}
For low-density models, this intermediate inclination range (neither pole-on nor edge-on) spans more than thirty degrees, with the range decreasing as the density increases.
{If such a feature is not considered in population studies, we may produce an observational bias in favor of extreme inclinations.}

The study of the dynamical grid offers valuable insights into the observed light curves. For light curves where a clear baseline is available, 
it is possible to make a rough estimate of the inclination angle, placing the object as either pole-on or edge-on, depending on the sign of the excess.
The challenge, however, as will be further discussed and illustrated below, lies in the fact that a clear baseline is not always identifiable. In many cases, variations in the disk's build-up and dissipation phases obscure the baseline, making it difficult to determine the diskless state directly (if even present in the lightcurve).
In such scenarios, the combined analysis of the light curve shape and the color-magnitude diagram (CMD) becomes crucial. 
These tools together can often provide the necessary information to classify an object as pole-on or edge-on, even in the absence of a well-defined baseline.

With this information, we can revisit Fig.~\ref{fig:examples_LCs} and incorporate CMDs as a diagnostic tool for inclination. The light curves in panels A and C correspond to pole-on stars, characterized by their positive excesses and reddening loops, while panel B represents an edge-on light curve, evident from the overall flux reduction due to disk absorption and little color variation.
However, distinguishing intermediate inclinations remains challenging. In such cases, the light curve and CMD may not provide definitive diagnostics, as seen in panel D. For this panel, the CMD suggests an ambiguous behavior.

\subsection{Diskless grid}\label{sec:model_disk-less}

For at least part of our sample, we expect to find lightcurves with a temporary diskless state (i.e., with a well-defined baseline; this subsample is defined in the next section).
By accounting for interstellar reddening, the corresponding CMD position can be used to estimate the stellar mass by comparing it with synthetic models of fast-spinning, diskless stars.
To support this goal, a grid of diskless stellar models was run, spanning 11 distinct stellar masses, 10 rotation rates, 10 inclination angles and 6 Main Sequence (MS) lifetimes ($\tau _{\rm ms}$), with the corresponding values shown in Table \ref{tab:models_disk-less_params}.
Note that $\tau _{\rm ms}$ is the fraction of time spent in the MS, having values between 0 and 1.
The methodology employed follows procedures similar to those outlined in the recent work of \citealt{Rubio2023MNRAS.526.3007R}. However, while their grid was limited to solar metallicities, our models incorporate Kurucz spectra \citep[for more details see][and references therein]{Rubio2023MNRAS.526.3007R} tailored for LMC and SMC metallicities.

\begin{table}
    \centering
    \begin{tabular}{lccccccc}\hline
        Mass & W & $\cos\,i$ & $\tau_{\rm ms}$ \\
        $(M_\odot)$ & & & \\\hline\hline
        1.7 & 0 & 0 & 0 \\
        2.0 & 0.33 & 0.11 & 0.40 \\
        2.5 & 0.47 & 0.22 & 0.65 \\
        3.0 & 0.57 & 0.33 & 0.85 \\
        4.0 & 0.66 & 0.45 & 1.00 \\
        5.0 & 0.74 & 0.55 & 1.25\\
        7.0 & 0.81 & 0.67 & \\
        9.0 & 0.87 & 0.78 & \\
        12.0 & 0.93 & 0.89 & \\
        15.0 & 0.99 & 1.00 & \\
        20.0 & & & \\\hline
    \end{tabular}
    \caption{Diskless model grid parameters. From left to right, the columns show the values for mass ($M$), rotation rates ($W$), inclination angle ($\mu = \cos i$) and Main Sequence life time ($\tau _{\rm ms}$), respectively.}
    \label{tab:models_disk-less_params}
\end{table}

\section{Data Analysis}\label{sec:data_analysis}

We now provide a detailed explanation of the processes applied to the raw sample defined in Sect.~\ref{sec:raw_sample}. These processes are designed to refine and categorize the data for subsequent analysis, following the necessary criteria. We outline each step in the pipeline, including data cleaning and filtering, and discuss the rationale behind the chosen methods. This systematic approach ensures that the raw sample is processed appropriately to achieve the objectives of this paper.

The core of our analysis is to classify lightcurves into different dynamical categories.
To achieve this, each light curve must satisfy specific criteria:
i) It must show variability above the defined detection threshold;
ii) We need to be able to identify the star's orientation (i.e., discriminate between pole-on and edge-on).

Targets that lack variability or 
{the CMD loop is not well sampled,} are removed,
as they do not meet both requirements.
Examples of stars with a lack of variability or with variability that does not allow us to diagnose orientation are those shown in panels E and D of Fig.~\ref{fig:examples_LCs}, respectively.
{Eclipsing binaries and/or ellipsoidal variables, which show strictly periodic signals with characteristic shapes (as illustrated by the lightcurve in panel F of Fig.~\ref{fig:examples_LCs}), were also removed}

At the end of this procedure, we obtain a sample of Be star candidates with clear signs of disk activity. Note that it is possible that the above criteria may exclude Be stars with persistent, stable disks that show no signs of variability during observations.
Table~\ref{tab:raw_sample_Summary} summarizes the results of this initial analysis. The first row shows the number of suspected eclipsing binaries, while the second lists the number of stars with no or unclear activity.
The third row indicates the number of targets that meet criteria i) and ii) and are, therefore, considered Be star candidates with disk activity.

\begin{table}
    \centering
    \caption{Summary of the filtering applied to the raw sample. The first row shows the number of suspected eclipsing binaries (S. E. Bin.),  the second the number of inactive light curves and those with unclear activity (Un. Act.) and the third the size of the sample of Be star candidates (Be Cand.). The total size of the raw sample is shown in the last row.
    }
    \begin{tabular}{lccc}
        Galaxy & SMC & LMC & Total \\ \hline \hline 
        S. E. Bin. & 4.5\% (44) & 2.8\% (59) & 3.3\% (103) \\ 
        Un. Act. & 34.0\% (336) & 41.2\% (884) & 38.9\% (1220) \\ 
        Be Cand. & 61.6\% (609) & 56.0\% (1201) & 57.8\% (1810) \\ 
        \hline 
        Total & 989 & 2144 & 3133 \\ 
    \end{tabular}
    \label{tab:raw_sample_Summary}
\end{table}

\begin{figure*}
    \centering
    \begin{tabular}{c}
        \includegraphics[width=.8\paperwidth]{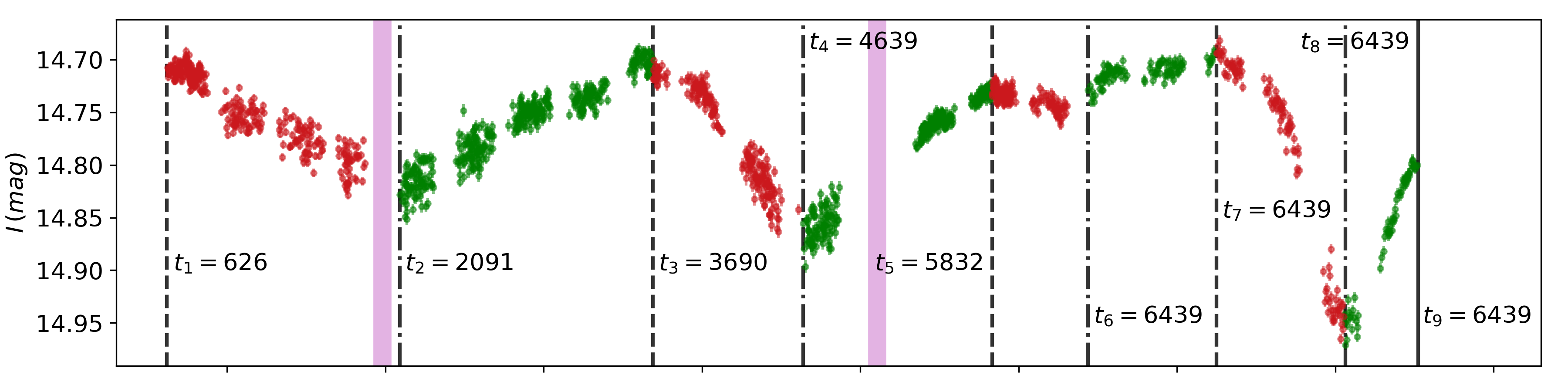} \\
        \includegraphics[width=.8\paperwidth]{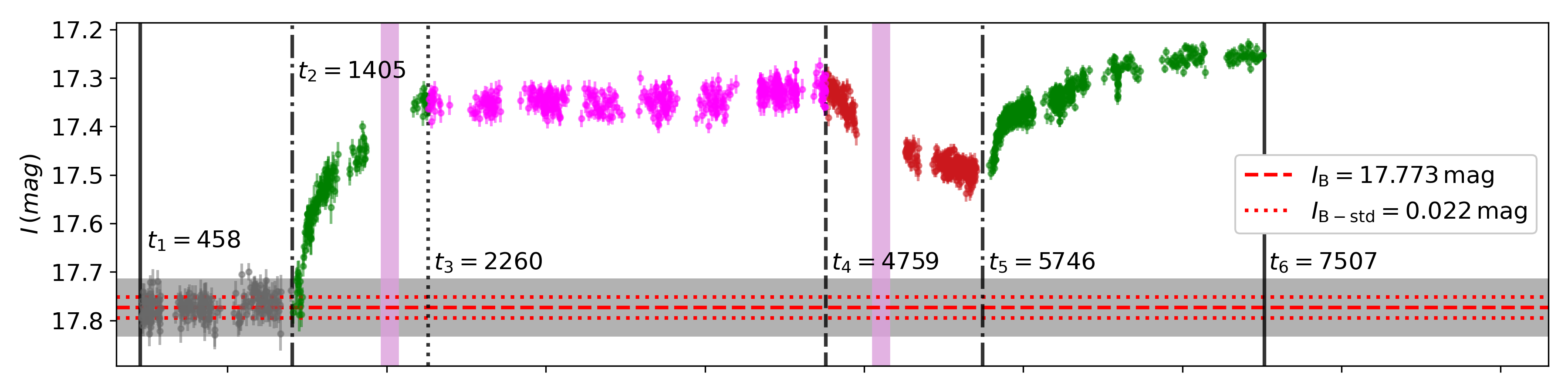} \\
        \includegraphics[width=.8\paperwidth]{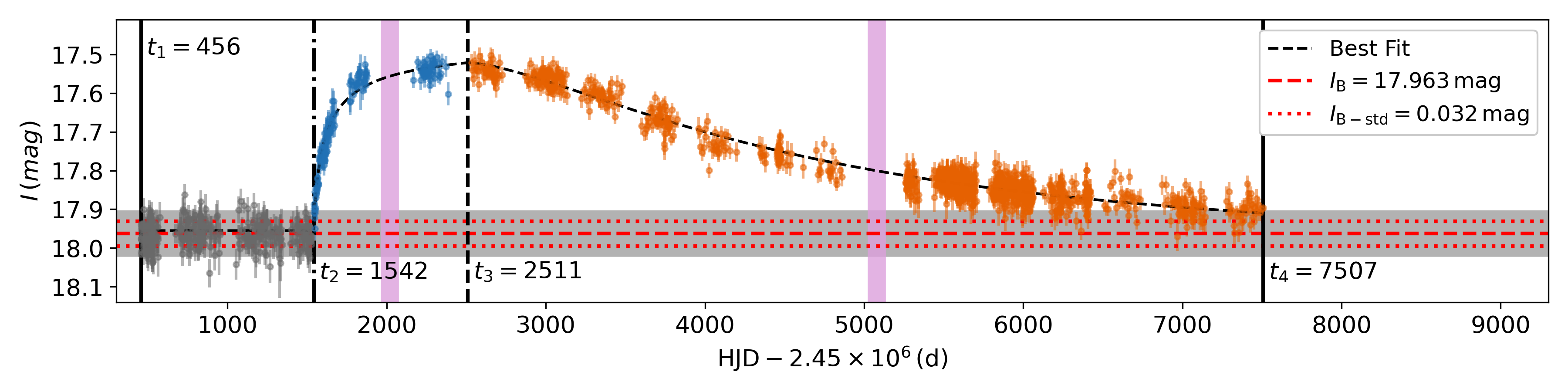} \\
    \end{tabular}
    \caption{Examples of lightcurves categorized according to the definitions in Sect.~\ref{sec:data_analysis}.
    The colors of data points and error bars represent different categories of variability: grey (1 -- baseline), blue (2 -- build-up), orange (3 -- dissipation), magenta (4 -- plateau), green (5 -- isolated build-up),  red (6 -- isolated dissipation).
    Horizontal red lines show the average baseline (dashed), and the respective standard deviation limits (doted), when available.
    Their respective values are shown in the legend.
    Vertical magenta bands shows the transition times among different observational phases.
    \textit{Top:} Lightcurve for target SMC\_SC2\_29957 for which the values of duty cycle and disk duty cycle estimated were $\rm DC = 0.55$ and $\rm DDC = 1.0$.
    \textit{Middle:} lightcurve for LMC\_SC10\_40703 with $\rm DC = 0.78$ and $\rm DDC =0.92$.
    \textit{Bottom:} lightcurve for LMC\_SC1\_164770, with $\rm DC =0.13$ and $\rm DDC = 0.83$. The dashed line shows the fit of Eqs.~\ref{eq:rimulo1} and \ref{eq:rimulo2}.
    Vertical black lines represents times marking the beginning and end of each category selected.
    %
    %
    }
    \label{fig:example_groups}
\end{figure*}

We now outline a methodology to classify the dynamic behavior of the sample of Be star candidates (for simplicity, we refer to this sample as the Be star sample, from now on).
To mitigate limitations due to visually inspecting the nearly twenty-year-long light curves, an interactive code was developed. This tool enables adjustment of the time scale under analysis, allowing detailed inspection of variations lasting from a few days to several months or years.
The behavior of each section of the lightcurve is then classified into seven categories, designed to cover all relevant dynamical scenarios. These categories,
illustrated in Fig.~\ref{fig:example_groups}, are defined as follows:

\begin{enumerate}
    \item Baseline (grey points in Fig.~\ref{fig:example_groups}): This phase corresponds to the photometric level where there is no disk and the mass loss is assumed to be either zero or negligible.

    \item Disk build-up (blue): 
    This phase corresponds to the growth of a new disk, or further growth of an older disk, due to enhanced mass loss by the star, leading to a monotonic increase or fading in brightness (depending on the inclination). This phase can only be identified if the light curve includes a baseline.
    
    \item Disk dissipation (orange): The disk-building phase is necessarily followed by a stage where mass ejection ceases, and the disk transitions into a dissipative state.

    \item Plateau (magenta): When a disk is fed for a sufficiently long period, it reaches maximum brightness and can remain in this state for an extended period \citep{haubois_2012}. 
    The negligible changes in brightness in this phase indicates a (nearly) constant mass loss rate.
   
    \item Isolated disk build-up (green): Similar to the disk-building phase, but not followed by disk dissipation.
   Such a phase may be followed by a plateau, as in the case of the middle panel of Fig.~\ref{fig:example_groups}, an isolated dissipation (top panel), or may be flagged as such due to lack of data.

    \item Isolated dissipation (red): Similar to the disk dissipation phase, but it is not preceded by a disk-building phase.
    Missing data also flags this option, as in the first panel of Fig.~\ref{fig:example_groups}.
    
    \item Unclassified (Purple): This category is used to designate sections of the light curve that could not be classified in any of the
    previous categories.
\end{enumerate}

The interactive code uses the aforementioned categories to estimate the duty cycle (DC), which is defined as the fraction of time a star spends 
{ejecting mass and feeding its disk.}
This parameter provides valuable insights into the temporal dynamics of mass ejection, helping to quantify how frequently and for how long a star exhibits disk-building activity over the observed period.
This quantity is defined as:
\begin{equation}\label{eq:duty_cycle}
    \mathrm{DC} = \frac{\Sigma t_{\dot{M}}}{t_{\rm tot} - t_{\rm unc}}\,, 
\end{equation}
where $\Sigma t_{\dot{M}}$ is the total time spent in phases with mass loss, including build-up, isolated build-up, and plateau, $t_{\rm tot}$ is the total observation time of the light curve and $t_{\rm unc}$ is the time span flagged as unclassified.

Another quantity of interest is the disk duty cycle, which represents the fraction of time a star harbors a disk (i.e., the fraction of time the star is observationally classified as a Be star). This is defined as:
\begin{equation}\label{eq:disk_duty_cycle}
    \mathrm{DDC} = \frac{\Sigma t_{\rm disk} }{ t_{\rm tot} - t_{\rm unc} } \,,
\end{equation}
where $\Sigma t_{\rm disk}$ is the total time spent in phases where a disk is present, which includes all categories except baseline and unclassified. 
{
We also estimate the outburst rate, $N_{\rm out}$, given by:
\begin{equation}
    N_{\rm out} = \frac{n_{\rm out}}{t_{\rm tot} - t_{\rm unc}}\,,
\end{equation}
where $n_{\rm out}$ is the total number of outbursts, indicating the number of times a lightcurve changes from a state where $\dot{M}=0$, or when it is negligible, to an active mass loss phase (such as categories 2, 4, and 5). 
}

Figure~\ref{fig:example_groups} illustrates the application of the above categories for three pole-on examples.
The top panel presents the lightcurve from target SMC\_SC2\_29957, which exhibits only isolated build-up and dissipation phases, maintaining a disk throughout the entire observing period. For this target, $\rm DC = 0.55$ and $\rm DDC = 1$.
LMC\_SC10\_40703 (middle panel) starts with a baseline, followed by a mass loss phase lasting approximately 5000\,d, which is then followed by isolated dissipation and build-up. For this star, $\rm DC=0.78$ and $\rm DDC=0.92$.
LMC\_SC1\_164770 (bottom panel) also starts with a baseline, followed by a disk build-up lasting approximately 1500\,d and a quite long dissipation (approximately 2500\,d). Here, $\rm DC=0.13$ and $\rm DDC=0.83$.
We also determine the outburst rate for each light curve: 
$N_{\rm out}=4/21.6=0.2\,{\rm yr}^{-1}$ (for SMC\_SC2\_29957), 
$N_{\rm out}=2/19.3=0.1\,{\rm yr}^{-1}$ (LMC\_SC10\_40703) and 
$N_{\rm out}=1/19.8=0.05\,{\rm yr}^{-1}$ (LMC\_SC1\_164770).


The bottom panel of Fig.~\ref{fig:example_groups} presents a light curve with a key property, relevant to this analysis: it features an \emph{isolated disk event}, characterized by an outburst preceded by a baseline and followed by complete dissipation, returning to baseline.
These disk events are ideal for studying mass loss and disk dynamics, as they are not influenced by the MRE 
\citep[see ][]{Rimulo_2018MNRAS.476.3555R}.
To better determine the characteristics of isolated disk events, {we adopted the formulae} derived by \citeauthor{Rimulo_2018MNRAS.476.3555R}:

\begin{equation}
    \Delta I_b(t) =  \Delta I_{\rm bu}^\infty \times \left( 1 - \frac{1}{1+ \left[ C_{\rm bu} (  t - t_{\rm B} )\right]^{\eta_{\rm bu}}}\, \right),\label{eq:rimulo1}
\end{equation}
{where $t_{\rm B} \leq t < t_{\rm D}$, for disk build-up and,}
\begin{equation}
    \Delta I_d(t) = \Delta I_b(t_{\rm D}) \times      \left( \frac{1}{1+ \left[ C_{\rm d} (  t - t_D )\right]^{\eta_{\rm d}}} \right),
    \label{eq:rimulo2}
\end{equation}
with $t \geq t_{\rm D}$ for dissipation.
Here, $\Delta I(t)$ is the amplitude of the photometric variation, $\Delta I_{bu}^\infty$ is the asymptotic value of $\Delta I(t)$, for the hypothetical case where mass loss does not cease and has a constant rate.
%
%
The rate coefficients ($C_{\rm bu}$ and $C_{\rm d}$) are free parameters that can assume different values for the build-up and dissipation phases.
$\eta_{\rm bu}$ and $\eta_{\rm d}$ were fixed, respectively, as $0.8$ and $1.4$ for the present work \citep[see][for more details]{Rimulo_2018MNRAS.476.3555R}.

Equations~\ref{eq:rimulo1} and \ref{eq:rimulo2} are fitted to isolated disk events.
An example is shown in the bottom part of Fig.~\ref{fig:example_groups}.
The best fit provides us with a photometric amplitude of $\Delta I = 0.423\,{\rm mag}$, a build-up phase  that begins at $t_{2}=1545\,{\rm d}$ lasting until $t_{3}=2511\,{\rm d}$, where dissipation takes place and ends at $t_{4}=7507\,{\rm d}$.
The build-up phase lasts for $t_{\rm b} =t_{3} -t_{2} = 969\,{\rm d}$, while dissipation occurs for ${t_{\rm d} =t_{4} -t_{3} = 4996\,{\rm d}}$.

\begin{figure}
    \centering
    \includegraphics[width=0.35\textwidth]{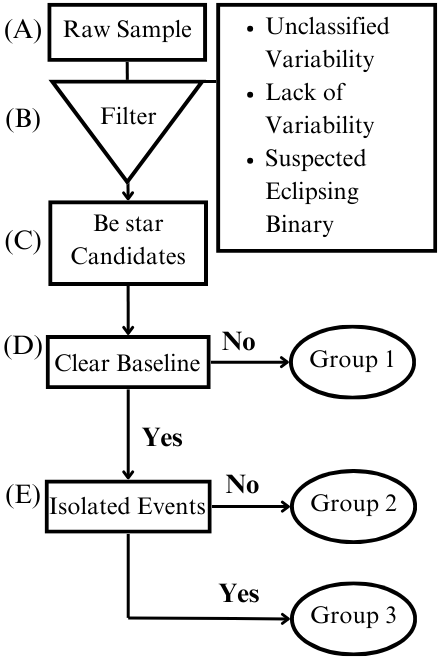}
    \caption{Schematic view of the samples defined in the paper. In the upper-right box we list the characteristics of the light curves that were removed to obtain the raw sample.}
    \label{fig:fluxogram_groups}
\end{figure}

%
This motivated us to categorize the Be star sample into three distinct groups with the following features:
\begin{itemize}
    \item {Group 1:} Lightcurves without a baseline phase. For this group, only the DC, DDC and $N_{\rm out}$ can be estimated. SMC\_SC2\_29957, in Fig.~\ref{fig:example_groups}, belongs to this group.
    \item {Group 2:}
    Targets, such as LMC\_SC10\_40703, from Fig.\ref{fig:example_groups}, with a baseline present, but without isolated disk events.
    This property allows for measuring the stellar mass, in addition to the above.
    \item {Group 3:} Light curves with isolated disk events (LMC\_SC1\_164770), which allow measuring the entire suite of quantities {outlined in this section}.
\end{itemize}
\noindent Table~\ref{tab:groupsmeasurements} summarizes the categories present and the quantities that can be extracted for each group.


A schematic view of our analysis procedure, starting from the raw sample and ending with the groupings described above, is shown in Fig.~\ref{fig:fluxogram_groups}.

\begin{table}[]
    \centering
    \caption{Measurements made for each group.}
    \begin{tabular}{cccc}\hline
        Quantity & G1 & G2 & G3 \\\hline\hline
        Baseline & No & Yes & Yes \\
        Complete Build-up & No & No & Yes \\
        Complete Dissipation & No & No & Yes \\
        Plateau & Yes & Yes & Yes \\
        Isolated Build-up & Yes & Yes & Yes \\
        Isolated Dissipation & Yes & Yes & Yes \\
        Unclassified & Yes & Yes & Yes \\\hline\hline
        DC & Yes & Yes & Yes \\
        DDC & Yes & Yes & Yes \\
        $N_{\rm out}$ & Yes & Yes & Yes \\
        $M$ & No & Yes & Yes \\
        $\Delta I$ & No & No & Yes \\
        $t_{\rm b}$ & No & No & Yes \\
        $t_{\rm d}$ & No & No & Yes \\\hline
    \end{tabular}
    \label{tab:groupsmeasurements}
\end{table}

\section{Results}\label{sec:results}

A general summary of the Be sample is presented in Table~\ref{tab:BeCandidatesSample}, which includes data for both Magellanic Clouds and the combined sample (last column). The top two rows summarize the inclination diagnostics.
The bottom two rows display, respectively, the numbers of stars that show baselines and those with events (either isolated or not).
%
The Be star sample consists of 1751 targets, predominantly composed of pole-on light curves (1639 stars, with 1095 from the LMC and 544 from the SMC), compared to 112 stars of the opposite orientation (59 from the LMC and 53 from the SMC).
There are two factors that contribute to the significant difference in the numbers for each orientation.
The first is that pole-on stars encompass a much large range of inclination angles, typically between 0 and approximately 60 degrees, while the edge-on stars generally have inclination angles larger than 80 degrees
(but this is rather uncertain and depends strongly on the density; see Fig.~\ref{fig:modelos_angulointermediario}).
This suggests that the number of pole-on stars should be around three to five times larger than that of edge-on stars.
The second factor is that the photometric signature of an outburst is more easily detected by a pole-on observer, since the associated photometric amplitudes are much larger in this case ( see Fig.~\ref{fig:modelos_angulointermediario}). 
%
One may also note that targets displaying baseline and isolated-event features are mostly LMC stars.
In Sect.~\ref{sec:discussion} we explore factors that possibly explain those differences.

The distribution of Be candidates among the three groups defined in Sect.~\ref{sec:data_analysis} and  Fig.~\ref{fig:fluxogram_groups} is detailed in Tab.~\ref{tab:groups}. The table also provides a breakdown based on different orientations.
The Be sample 
includes 935 group 1 stars, 464 group 2 stars, and 352 stars in group 3.
Recall that group 1 stars are those with complex light curves, lacking a simple pattern of disk formation and dissipation, and without a baseline. These constitute 41\% of the stars in the LMC and the majority of the stars in the SMC (78\%).
The number of stars decreases in groups 2 and 3 for both galaxies, as expected, since the light curves in these groups must meet progressively stricter criteria.


In Sect.~\ref{sec:results_wholesample}, all three groups are combined to study how the entire Be sample behaves in terms of outburst rate, duty cycle, and disk duty cycle. 
The following section, Sect.~\ref{sec:results_Mass}, presents the estimated stellar masses for groups 2 and 3, comparing these results with findings from the literature.
Using the mass information, we explore how this parameter correlates with DC, DDC, and \( N_{\rm out} \) in Sect.~\ref{sec:results_mass_dependency}. 
Finally, the last subsection focuses on the results for lightcurves possessing isolated disk events (Group 3).


\begin{table}
    \centering
    \caption{
    Sample size of variable Be stars identified according to orientation (top two rows) and  activity patterns (bottom two rows).
    }
    \begin{tabular}{lccc}
        Galaxy & LMC & SMC & Total\\ \hline 
        Pole-on & 1095 & 544 & 1639\\ 
        Edge-on & 59 & 53 & 112\\ 
        Baseline & 616 & 132 & 748\\ 
        Isolated Events  & 219 & 39 & 258\\ 
    \end{tabular}
    \label{tab:BeCandidatesSample}
\end{table}

\begin{table}
    \centering
    \caption{Summary of the target distribution across each group, divided by different metallicities (columns) and orientations (rows).
    The numbers in bold indicate the total number of light curves in each group.}
    \begin{tabular}{clccc}
        Group & Orientation & LMC & SMC & Total\\\hline
        1 & Pole-on & 427 & 421 & 848 \\
         & Edge-on & 43 & 44 & 87\\
         & Total & 470 & 465 & \textbf{935}\\\hline
        2 & Pole-on & 379 & 71 & 450\\
         & Edge-on & 8 & 6 & 14\\
         & Total & 387 & 77 & \textbf{464}\\\hline
        3 & Pole-on & 289 & 52 & 341\\
         & Edge-on & 8 & 3 & 11\\
         & Total & 297 & 55 & \textbf{352}
    \end{tabular}
    \label{tab:groups}
\end{table}

\subsection{Outburst rate, DC and DDC}\label{sec:results_wholesample}

\begin{figure}
    \centering
    \begin{tabular}{c}
        \includegraphics[width=0.95\linewidth]{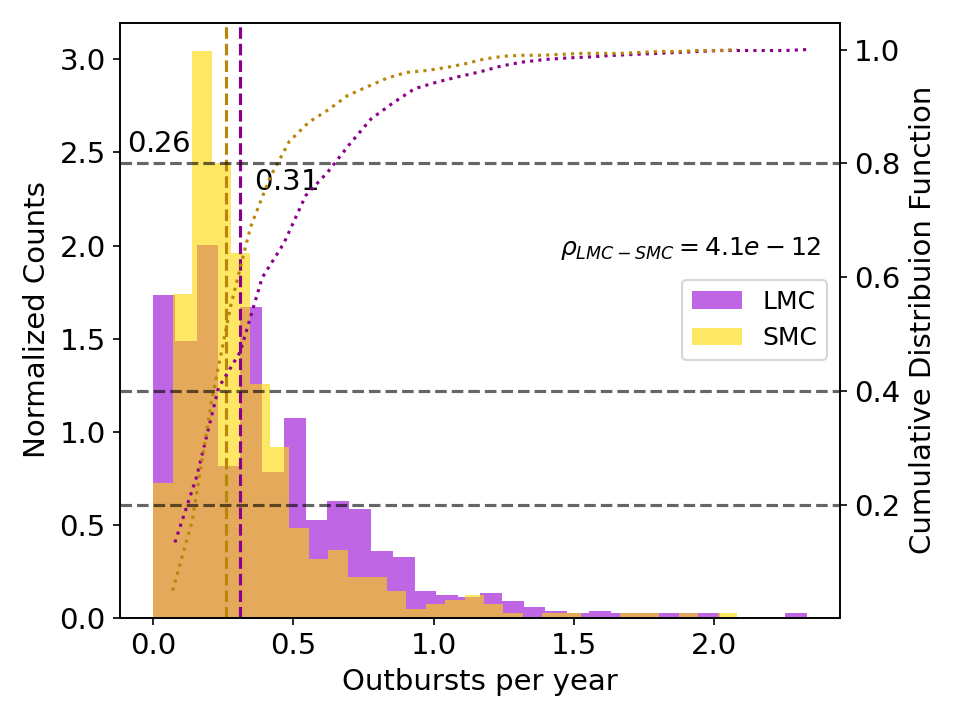}\\
    \end{tabular}
    \caption{Normalized histogram of the number of outburst per year ($N_{\rm out}$) for LMC (purple) and SMC (yellow) stars of the entire Be star sample. The numbers of light curves of each sample are shown in the legend: 1154 for LMC and 597 SMC lightcurves. 
    %
    Median values are shown by vertical dashed lines.
    The cumulative distribution functions, assisted by horizontal dashed line for 20, 40 and 80\%, are represented by dotted lines (right-hand axis).
    The value of the Kolgomorov-Smirnov test applied to both samples is $\rho_{\rm LMC - SMC}$).}
    \label{fig:hist_Nout}
\end{figure}

The outburst rate for the entire Be sample is presented in Fig.~\ref{fig:hist_Nout}, where LMC and SMC targets are represented by purple and yellow histograms, respectively.
Due to the imbalance in the number of Be candidates between galaxies, with approximately {66\% ({1154}, from a total of 1751 light curves)} of the targets located in the LMC, the histograms in Fig.~\ref{fig:hist_Nout} are plotted with normalized counts. This ensures that the histograms cover the same area, enabling a more effective comparison.
The {SMC} distribution exhibits a well-defined peak near $0.2\,{\rm yr^{-1}}$, with a corresponding median (indicated by the dark yellow vertical dashed line) of $0.26\,{\rm yr^{-1}}$.
The LMC stars show a not too dissimilar distribution and median but lack a clearly defined peak.

To test whether both samples originate from the same underlying distribution or are statistically distinct, a Kolmogorov--Smirnov (KS) test was performed. The resulting $p$-value of the KS test, $\rho_{\rm LMC-SMC} = 4.1 \times 10^{-12}$, allows us to reject the null hypothesis, indicating that the samples are statistically distinct.
This is further corroborated by the cumulative distribution function (right-hand axis), which reveals that the SMC population typically exhibits lower values of $N_{\rm out}$. 

\begin{figure*}
    \centering
    \begin{tabular}{cc}
        \includegraphics[width=.49\textwidth]{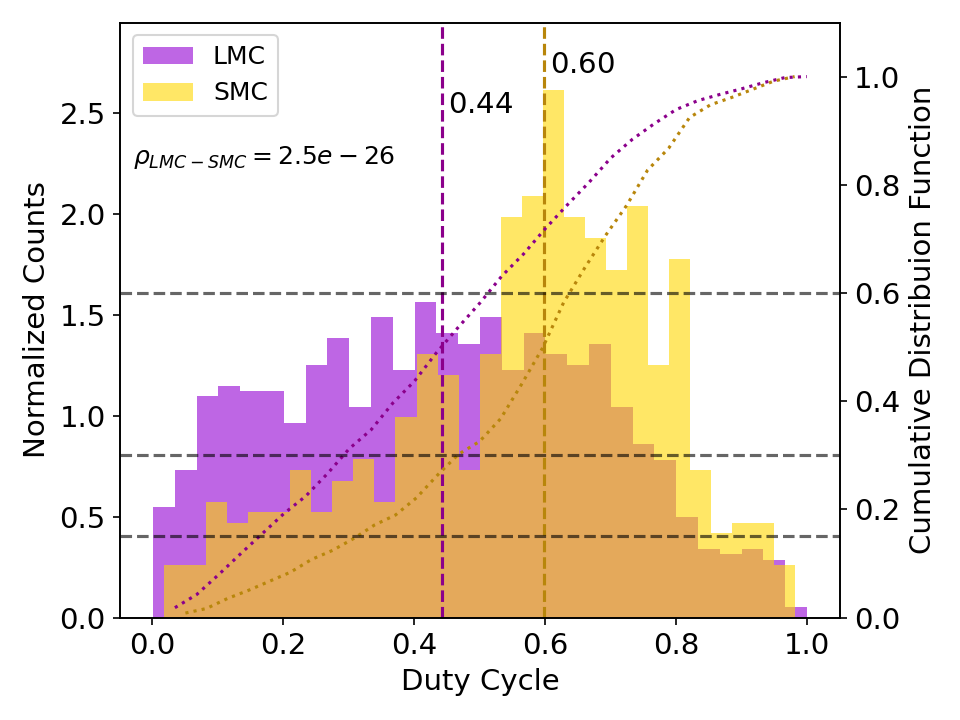} & \includegraphics[width=.49\textwidth]{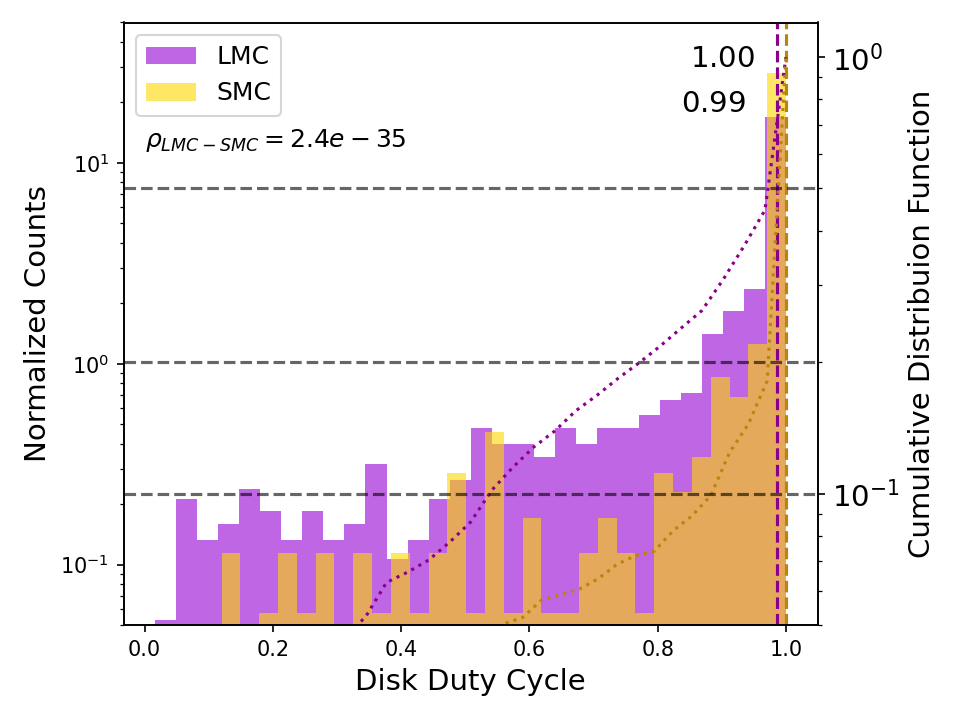} \\
    \end{tabular}
    \caption{Same as Fig.~\ref{fig:hist_Nout} for the DC (left panel) and DDC (right panel). 
    }
    \label{fig:hist_dutycycle}
\end{figure*}

The DC normalized histogram is shown in the left panel of Fig.~\ref{fig:hist_dutycycle}.
When comparing both populations, it is clear that the SMC sample is comprised of more active stars, with a broad peak around $\rm DC=0.6$. The LMC, on the other hand, exhibits an even broader spread centered around $\rm DC\sim0.4$.
This trend is also clearly reflected in the CDF: the {LMC} values grow almost linearly, while the {SMC} shows slower initial growth followed by a rapid increase later.
These populations are also statistically distinct, as confirmed by the KS test.

The right panel of Fig.~\ref{fig:hist_dutycycle} shows the DDC histogram. 
Note that, given the large dynamical range of this parameter, the ordinates are in log scale.
The respective median values of $0.99$ (LMC) and $1.00$ (SMC) indicate that the vast majority of stars in both samples keep their disk for most of the time.
An example of a star that maintained its disk during the entirety of observations is SMC\_SC2\_29957 (Fig.~\ref{fig:example_groups}.)
Despite the similarities in the median values, the cumulative distribution function shows that the SMC sample is more strongly peaked towards higher values of DDC.
Those differences in behavior are also corroborated by the KS test.

\subsection{Mass determination}
\label{sec:results_Mass}

\begin{figure*}
    \begin{tabular}{c c}
        \includegraphics[width=0.45\textwidth]{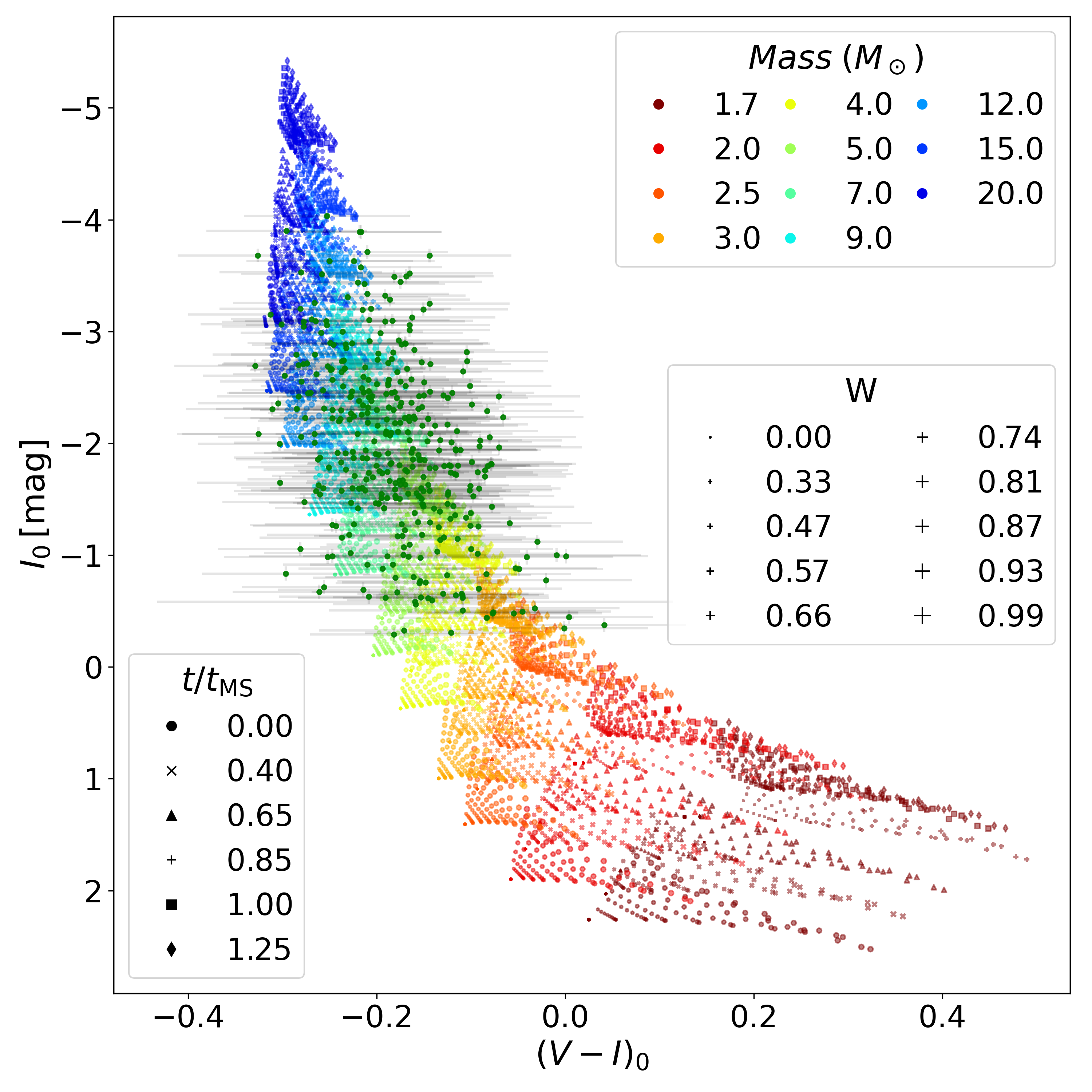} & \includegraphics[width=0.45\textwidth]{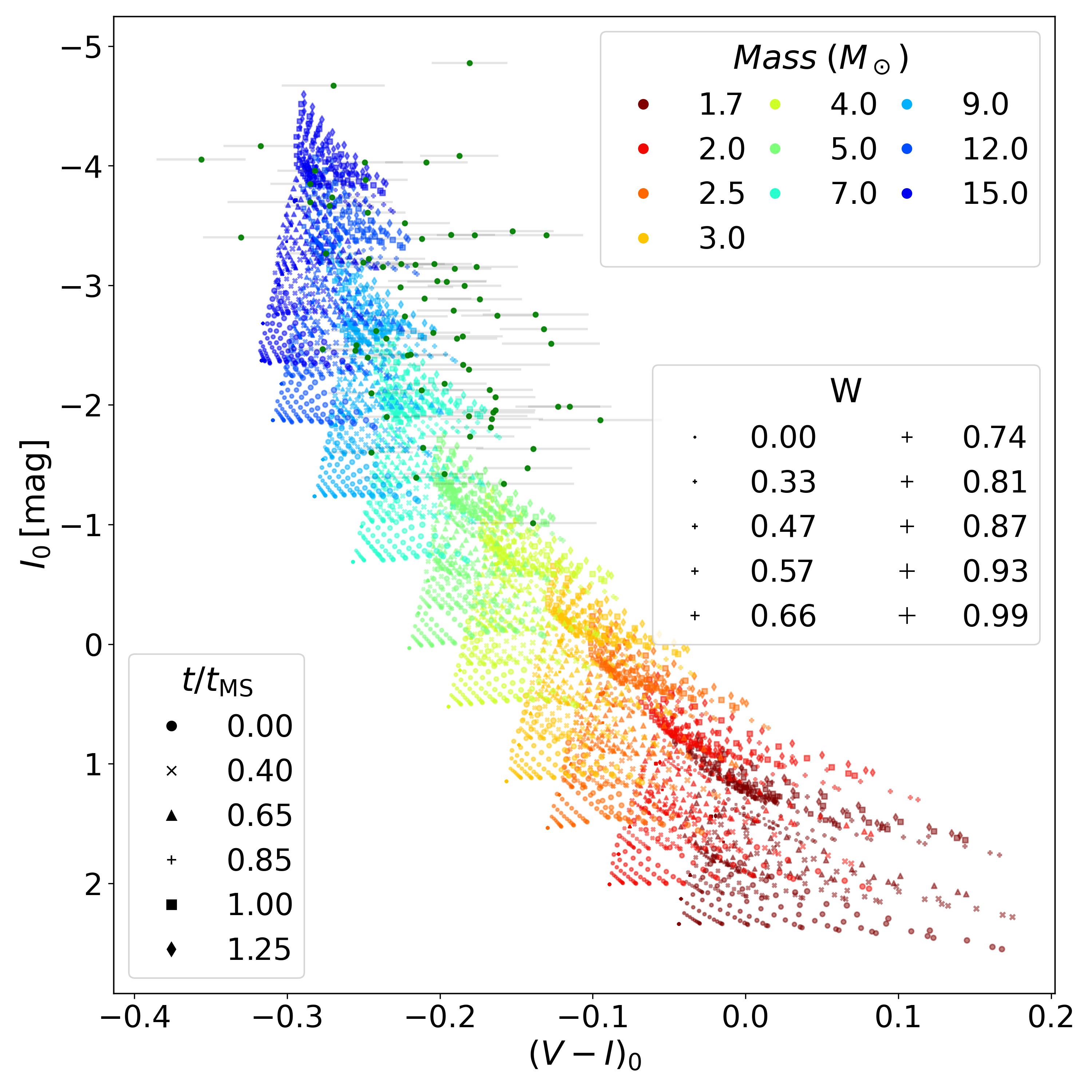}\\
    \end{tabular}
    \caption{Color–magnitude diagram of simulated diskless stars (Table \ref{tab:models_disk-less_params}) for LMC (left) and SMC (right).
    The colors ranging from dark red to dark blue correspond to the eleven values of $M_\odot$, as shown in the upper right legend.
    The ten rotation values are represented by symbols sizes (middle right legend).
    The different symbols represents the stellar Main Sequence age (bottom left legend).
    The data for stars of groups 2 and 3 are represented by the green dots with error bars.
    }
    \label{fig:cmd_disk-less}
\end{figure*}

We use the baseline magnitudes of groups 2 and 3, unreddened according to the procedure outlined in Sect.~\ref{sec:raw_sample}. The $E(V-I)_{\rm median}$ values of \citet{Skowron2021ApJS..252...23S} were adopted, and the distance moduli used were $18.49\pm0.06\,{\rm mag}$ \citep{Pietrzynski_2013Natur.495...76P} and $18.977\pm0.016\,{\rm mag}$ \citep{Graczyk2020ApJ...904...13G} for the LMC and SMC, respectively.

Figure~\ref{fig:cmd_disk-less} compares models (colored markers) and the corresponding dereddened baseline CMD positions (green dots with error bars).
There is a general agreement between the theoretical main sequence strip and the data, within the errors; however, the agreement is better for LMC stars than for SMC ones.
For the LMC, the CMD indicates that groups 2 and 3 consist of stars with masses ranging from approximately 3 to $20\,(M_\odot)$, falling within the expected range for the Be phenomenon \citep{Rivinius2013A&ARv..21...69R}.
In contrast, both the upper and lower mass limits for SMC stars are higher than those observed for LMC stars; however, the SMC sample size is significantly smaller (Tab.~\ref{tab:groups}).

It is well known (e.g., \citealt{Georgy_2013A&A...553A..24G}) that the effects of mass, rotation rate, and age on a star's position in the CMD are degenerate, as evidenced by the significant overlap of points in the theoretical CMDs shown in Fig.~\ref{fig:cmd_disk-less}. To address this degeneracy and obtain a mass estimate for each star, we utilize the Markov Chain Monte Carlo (MCMC) Python implementation, \textsc{emcee} {\citep{Foreman-Mackey_2013PASP..125..306F}.
As the likelihood function we use the $\chi^2$ defined by \cite{Robitaille2007ApJS..169..328R} (their Eq. 6), given by:
\begin{equation}
\chi^2 = \sum_{X} \sum_{i=1}^N \left[ \frac{\log( X_{\rm obs,\,\textit{i}}/X_{\rm mod,\,\textit{i}} )}{\sigma_{\rm obs,\, \textit{i}}}
 \right] ^2 \,, 
\end{equation}
{where $X_{\rm obs,\,\textit{i}}$ are the baseline magnitudes in the photometric filter $X$ ($B$, $V$ and $I$), $\sigma_{\rm obs,\,\textit{i}}$ are the respective errors and $X_{\rm mod,\,\textit{i}}$}
represents the photospheric models (Sect.~\ref{sec:model_disk-less}).
}
{It is important to note that among the 816 stars in Groups 2 and 3, $V$-band data are available for 468 light curves (389 from the LMC and 79 from the SMC), while $B$-band data were obtained for 50 targets (28 from the LMC and 22 from the SMC). 
We adopted, as a minimum requirement, the availability of data in two bandpasses to perform the MCMC simulation. Consequently, the stellar mass was determined for 468 stars within Groups 2 and 3.}

\begin{figure}
    \centering
    \includegraphics[width=0.49\textwidth]{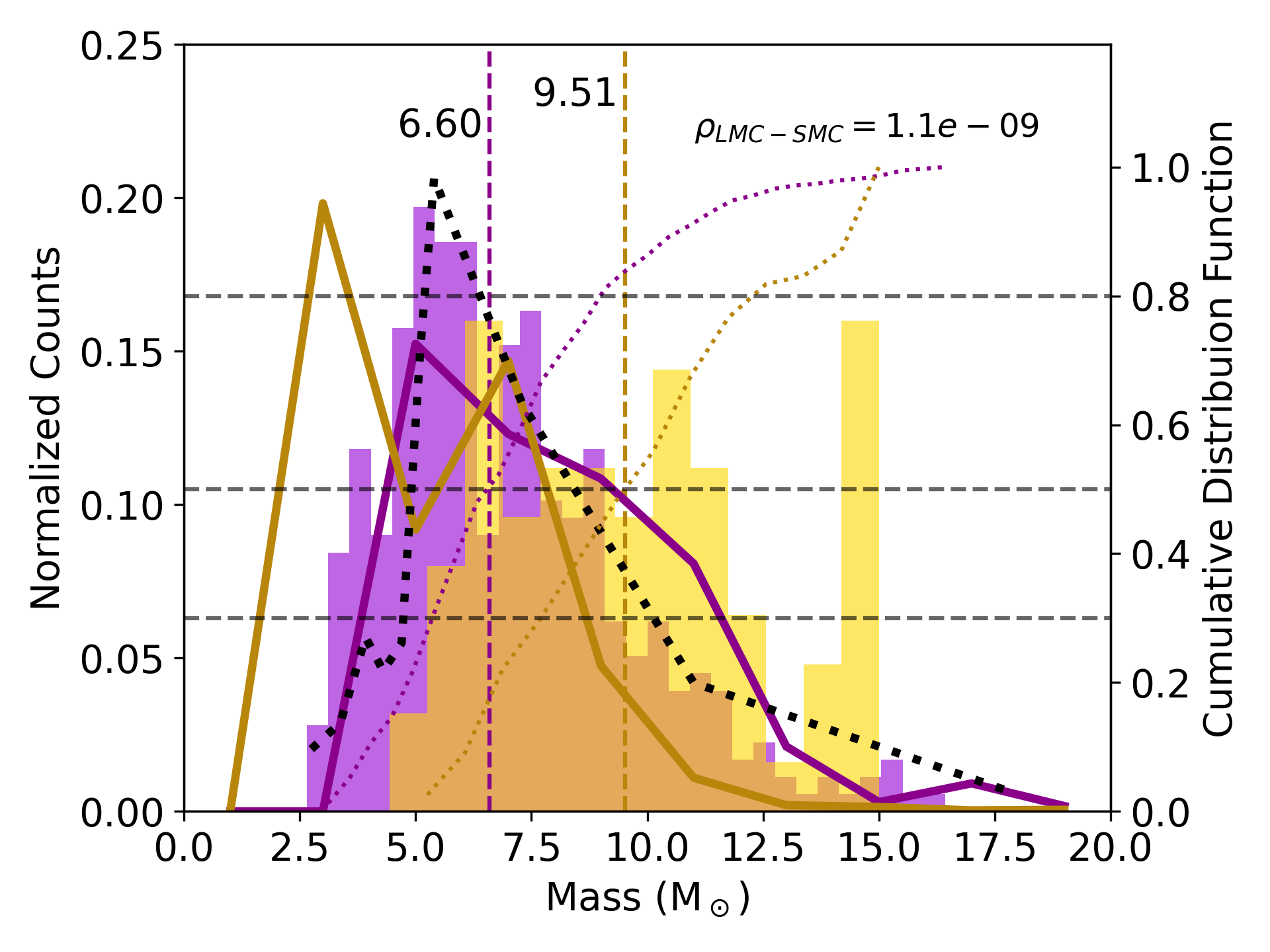}
    \caption{
    Similar to Fig.~\ref{fig:hist_Nout} for the masses obtained for 468 stars of groups 2 and 3\ (LMC population in purple and yellow represents SMC targets).
    %
    The expected Be stars {distribution} estimated from the literature (see text) are shown by thick lines. The purple line uses data from \citet{Martayan2006A&A...452..273M}, the yellow from \citet{Martayan_2007A&A...462..683M} and the dotted black line was extracted from \citet{Navarete2024ApJ...970..113N}.
    %
    }
    \label{fig:Mass_hist}
\end{figure}

The results of the MCMC simulations are shown in Fig.~\ref{fig:Mass_hist}. 
The mass distributions of LMC and SMC stars are quite distinct: while the higher-metallicity population has a peak near $5\,M_\odot$, the SMC distribution has a flatter shape and presents two peaks, one near $6.5$ and the second close to $14.5\,M_\odot$.
The contrast between samples is evident in their respective median values, $6.60$ for LMC and $9.51\,M_\odot$ for the SMC, as well as in the markedly different shapes of their CDFs.
Furthermore, the KS test indicates that the probability of both samples originating from the same underlying distribution is highly unlikely.


For comparison, we plot in Fig.~\ref{fig:Mass_hist} the expected Be star mass distribution for the LMC and SMC,
obtained by {multiplying} the IMF of \citet{Kroupa_2001MNRAS.322..231K}
with the Be/(B+Be) fractions estimated from different sources.
Using the data of Fig.~9 of \citet{Martayan2006A&A...452..273M} (purple line), we produce a mass distribution of LMC stars.
Utilizing Fig.~6 of \citet{Martayan_2007A&A...462..683M} (yellow line), an estimate of the Be star mass distribution for the SMC is obtained.
Lastly, from the data of Fig.~16 of \citet{Navarete2024ApJ...970..113N} (black dotted line), this distribution was obtained for the young stellar cluster NGC\,0330.
%
%
The results are somewhat ambiguous to interpret. The mass distribution for the LMC presented by \citeauthor{Martayan2006A&A...452..273M} closely resembles that of our sample, with both exhibiting a peak at a similar mass.
However, the other distributions obtained for the SMC (yellow and dotted black lines in Fig.~\ref{fig:Mass_hist}) differ significantly from the mass distribution of our SMC sample. 
The most striking differences are the excess of high-mass Be stars and the noticeable scarcity of low-mass stars in our sample.

\subsection{Mass dependency}\label{sec:results_mass_dependency}

\begin{figure}
    \centering
    \begin{tabular}{c}
        \includegraphics[width=0.99\linewidth]{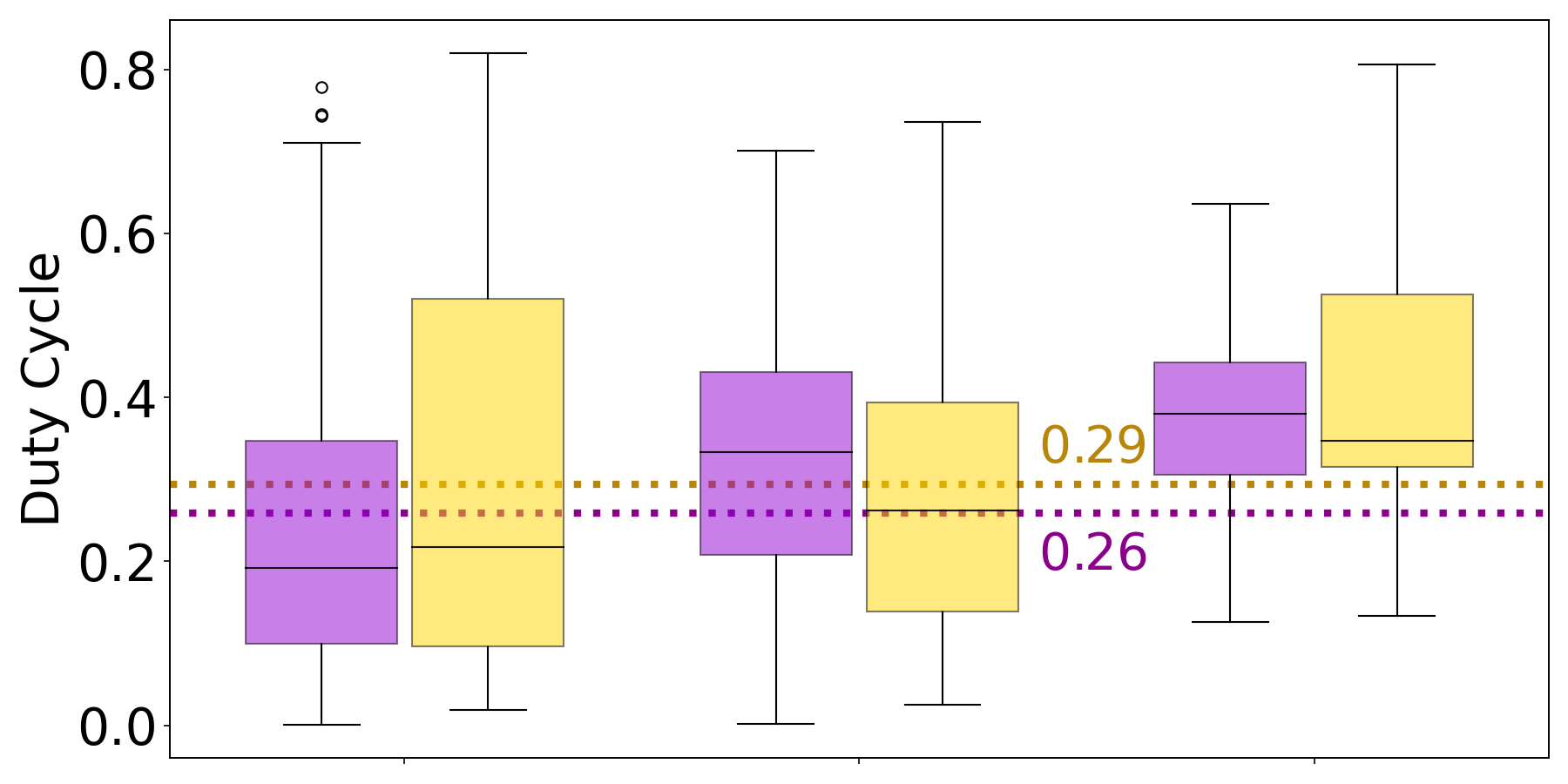}\\
        \includegraphics[width=0.99\linewidth]{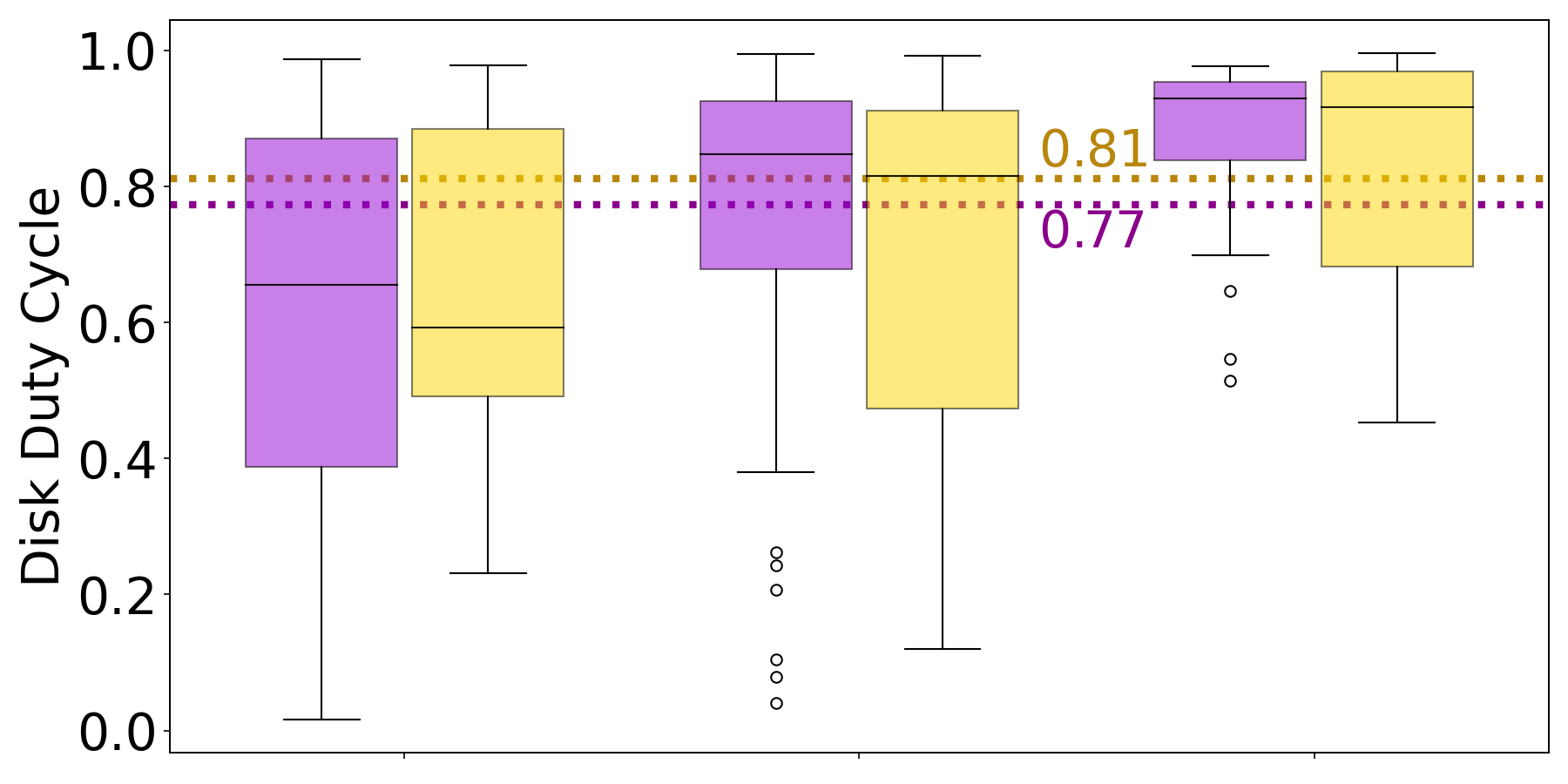}\\
        \includegraphics[width=0.99\linewidth]{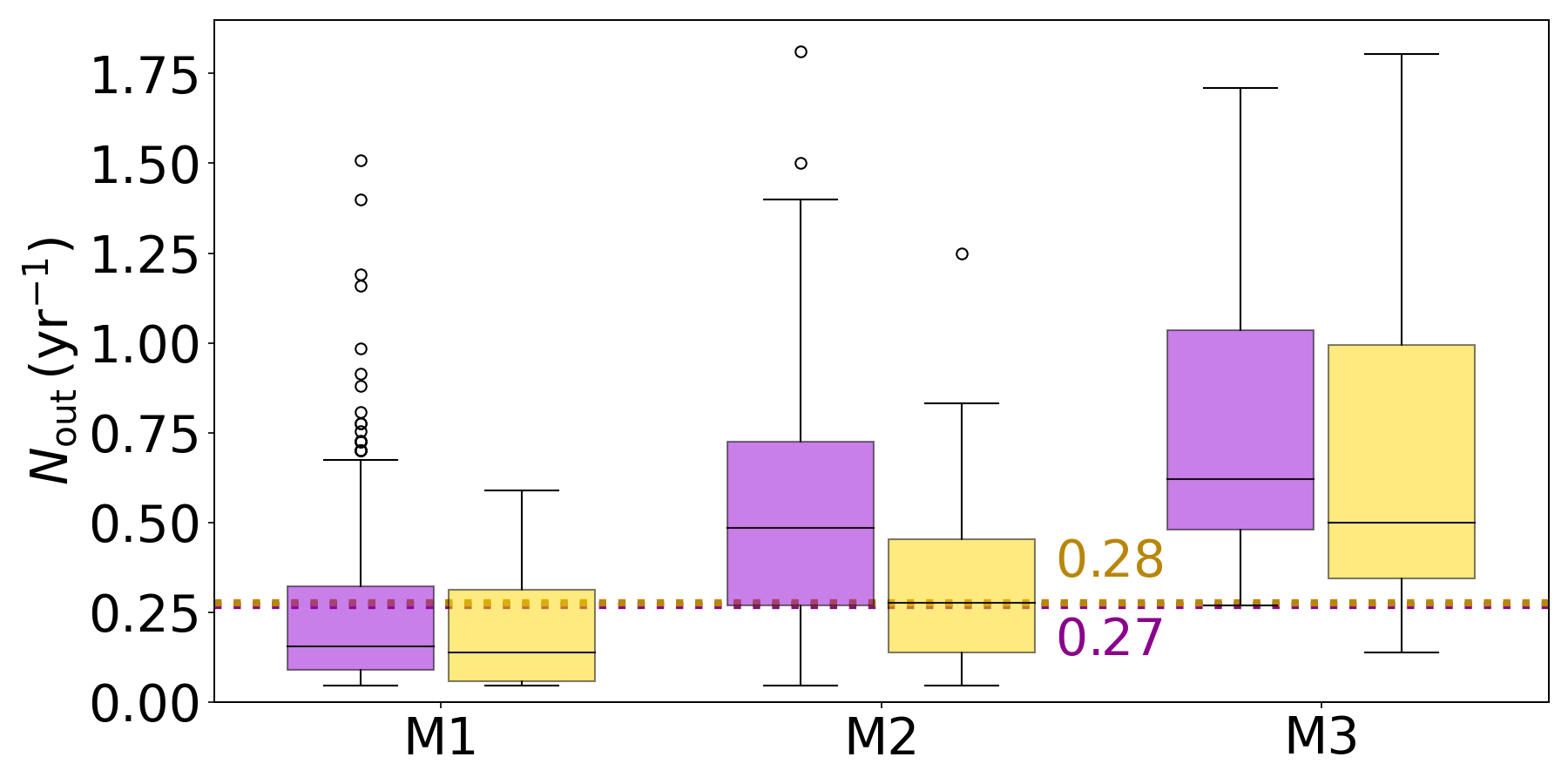}
    \end{tabular}
    \caption{
    Boxplots for the DC (top), DDC (middle) and $N_{\rm out}$ (bottom), combining the subsample that exhibits a baseline (groups 2 and 3), separated into three mass intervals (see text) for both galaxies {(purple for LMC and yellow for SMC stars).}
    Box plots summarize the distribution of the data, with the box indicating the interquartile range (IQR) and the line inside the box representing the median. The whiskers extend to 1.5 times the IQR.
    {The dotted lines represent the median values of the entire sample, with all mass bins combined.}
    The Kolmogorov-Smirnov test was run to compare the distributions of each mass bin from different metallicities, resulting
    $\rho_{\rm M_{\rm 1}}=0.36$,
    $\rho_{\rm M_{\rm 2}}=0.36$,
    $\rho_{\rm M_{\rm 3}}=0.21$ (for DC),
    $\rho_{\rm M_{\rm 1}}=0.81$,
    $\rho_{\rm M_{\rm 2}}=0.47$,
    $\rho_{\rm M_{\rm 3}}=0.61$ (for DDC),
    $\rho_{\rm M_{\rm 1}}=0.25$,
    $\rho_{\rm M_{\rm 2}}=1.5\times10^{-4}$ and
    $\rho_{\rm M_{\rm 3}}=0.42$ (for $N_{\rm out}$).
    }
    \label{fig:boxplot_group2and3}
\end{figure}

The mass values estimated in the last section are now used to study whether this parameter affects the variability diagnostics (DC, DDC and $N_{\rm out}$) derived above.
To address the limitations of our samples, we divide the stars into three mass ranges: $M_1$ corresponds to stars with ${M < 7\, M_\odot}$, $M_2$ includes stars with ${7\,M_\odot < M < 12\, M_\odot}$ and $M_3$ comprises stars that have ${M>12 \,M_\odot}$.
{{The 468 stars for which stellar mass was determined are distributed among the mass bins as follows: 217, 151, 20 (LMC) 19, 44 and 17, in increasing order of mass.}}

The mass dependency is shown in Fig.~\ref{fig:boxplot_group2and3}, where box plots of the DC (top) and DDC (middle) and $N_{\rm out}$ (bottom) are shown for each mass bin defined above, for both galaxies.
A box plot is used here to visually summarize the distribution of a dataset by highlighting key statistics. Its components are: 
i)\,Box: represents the interquartile range (IQR), covering the middle 50\% of the data, from the 25th percentile (Q1) to the 75th percentile (Q3). The line inside the box marks the median (50th percentile).
ii)\,Whiskers: Extend from the edges of the box to indicate the range of the data. They typically span up to 1.5 times the IQR beyond Q1 and Q3.
iii)\,Outliers: Data points that fall outside the whiskers are considered outliers and are displayed as individual markers (circles).

The results show a clear correlation between mass and DC for both galaxies. 
For the LMC, the lowest mass bin have a median values of ${ \left< \rm DC (M_{\rm 1})\right> }\simeq 0.2$ and this value grow with mass, reaching ${ \left< \rm DC (M_3)\right> } \sim 0.4$ for the highest mass bin.
A similar behavior is seen for the SMC.

Similar to the DC, a correlation with mass is also observed for the DDC in both galaxies (middle plot of Fig.~\ref{fig:boxplot_group2and3}), indicating that massive stars tend to harbor disks for longer than their less bright siblings.
As with the DC, comparing results within the same mass interval across different metallicities yields similar outcomes. 
One notable difference between the samples from each galaxy is the spread of the data, which is systematically larger for the SMC, except in the case of the DDC for $M_1$.

Finally, the bottom plot of Fig.~\ref{fig:boxplot_group2and3} a similar trend, with a marked increase in the number of outbursts per year with mass. 
Typically, $M_{\rm 3}$ stars experience an outburst rate that is three times higher than that of $M_{\rm 1}$ stars. For $M_{\rm 1}$, the outburst rate is approximately 0.2 events per year, increasing to around 0.6 events per year for $M_{\rm 3}$.


For the three variability diagnostics analyzed in this section, a comparison of distributions within the same mass bin across the two metallicities reveals that, according to the KS test, they are statistically indistinguishable, except for the outburst rate samples in the $M_{\rm 2}$ category (see numbers in the caption of Fig.~\ref{fig:boxplot_group2and3}).

\subsection{Group 3: parameters of isolated events and their mass dependence}\label{sec:result_isolatedevents}



\begin{figure}
    \centering
    \begin{tabular}{c}
        \includegraphics[width=0.45\textwidth]{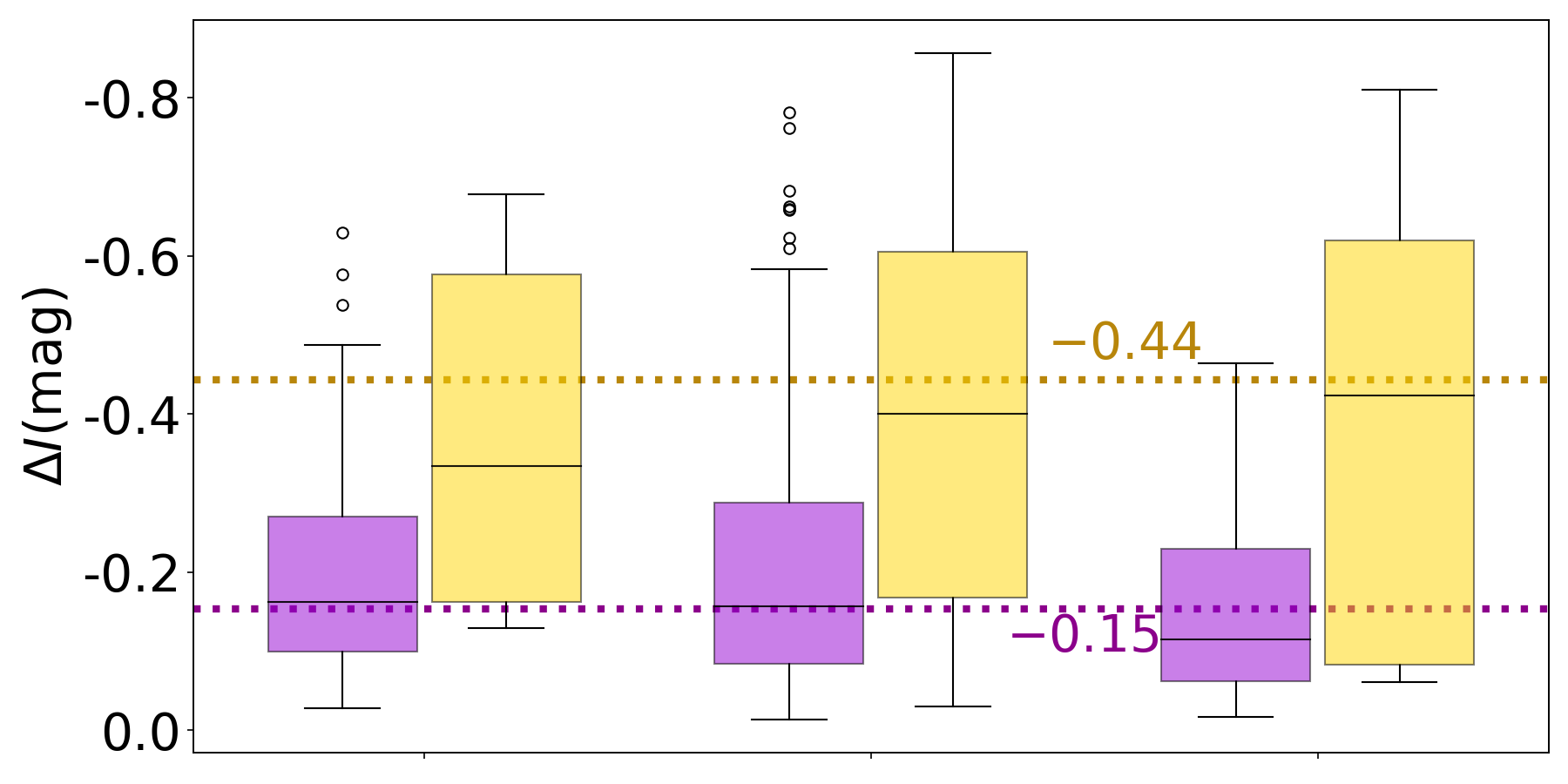} \\
        \includegraphics[width=0.45\textwidth]{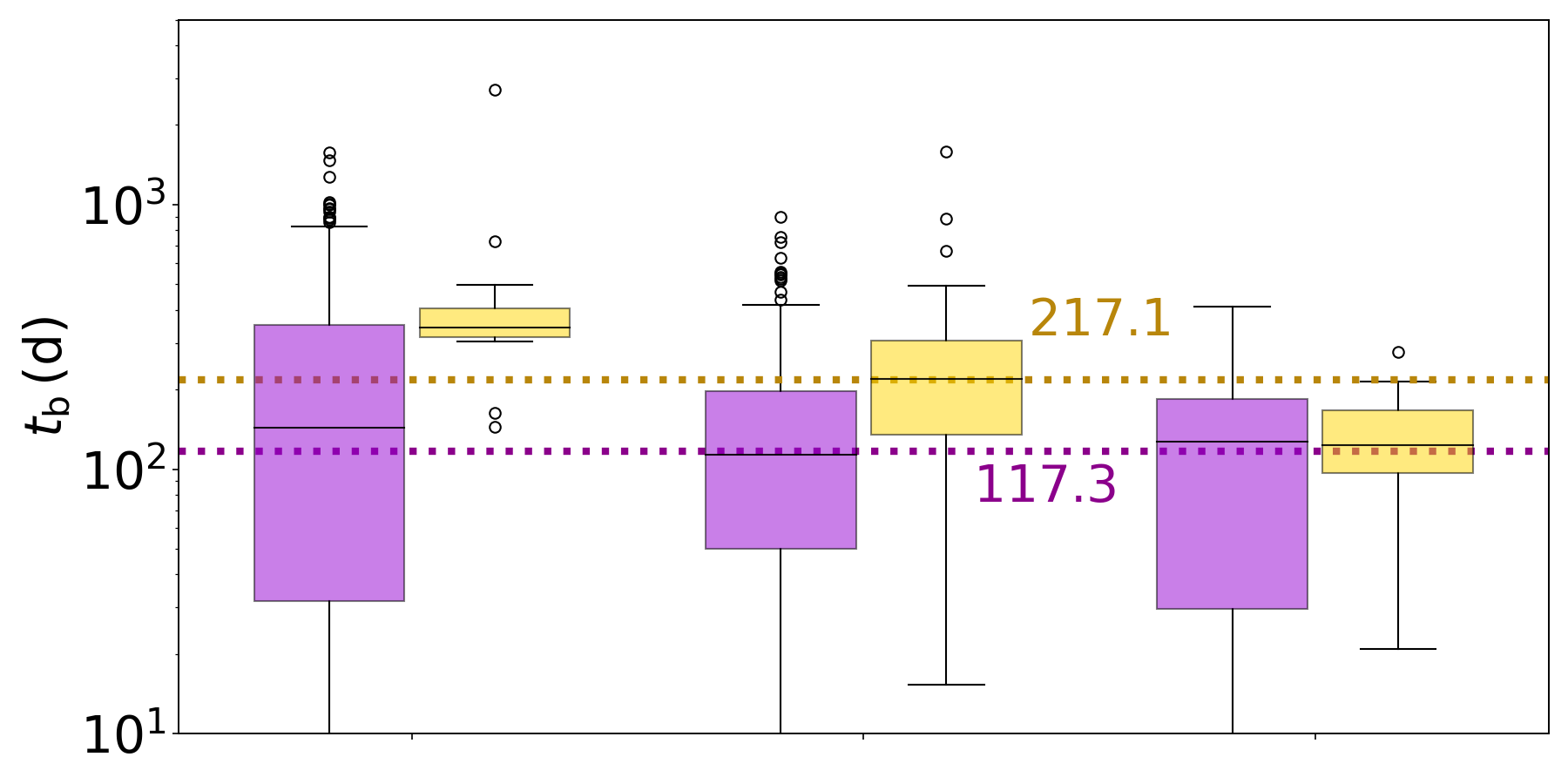} \\
        \includegraphics[width=0.45\textwidth]{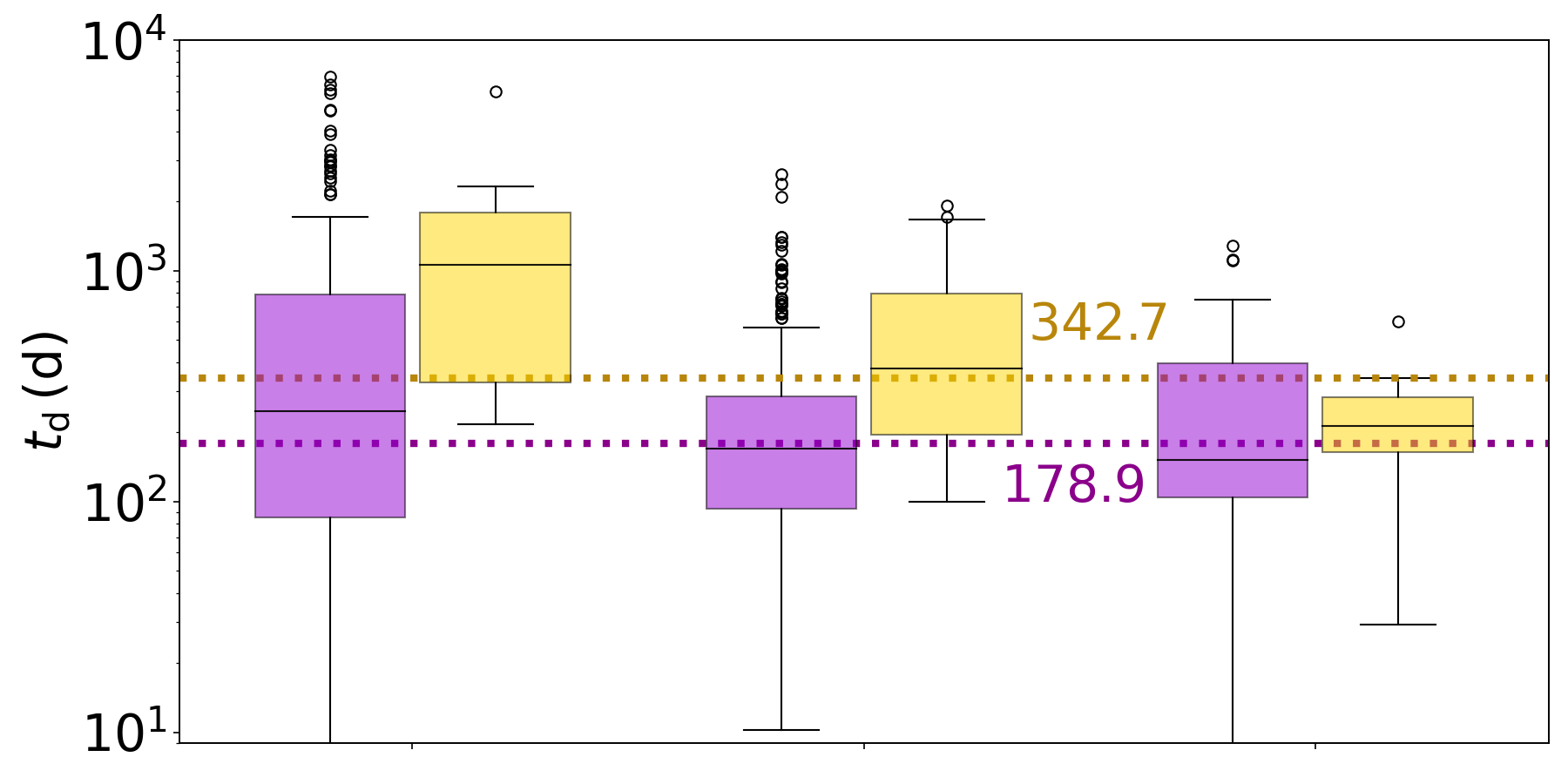} \\
        \includegraphics[width=0.45\textwidth]{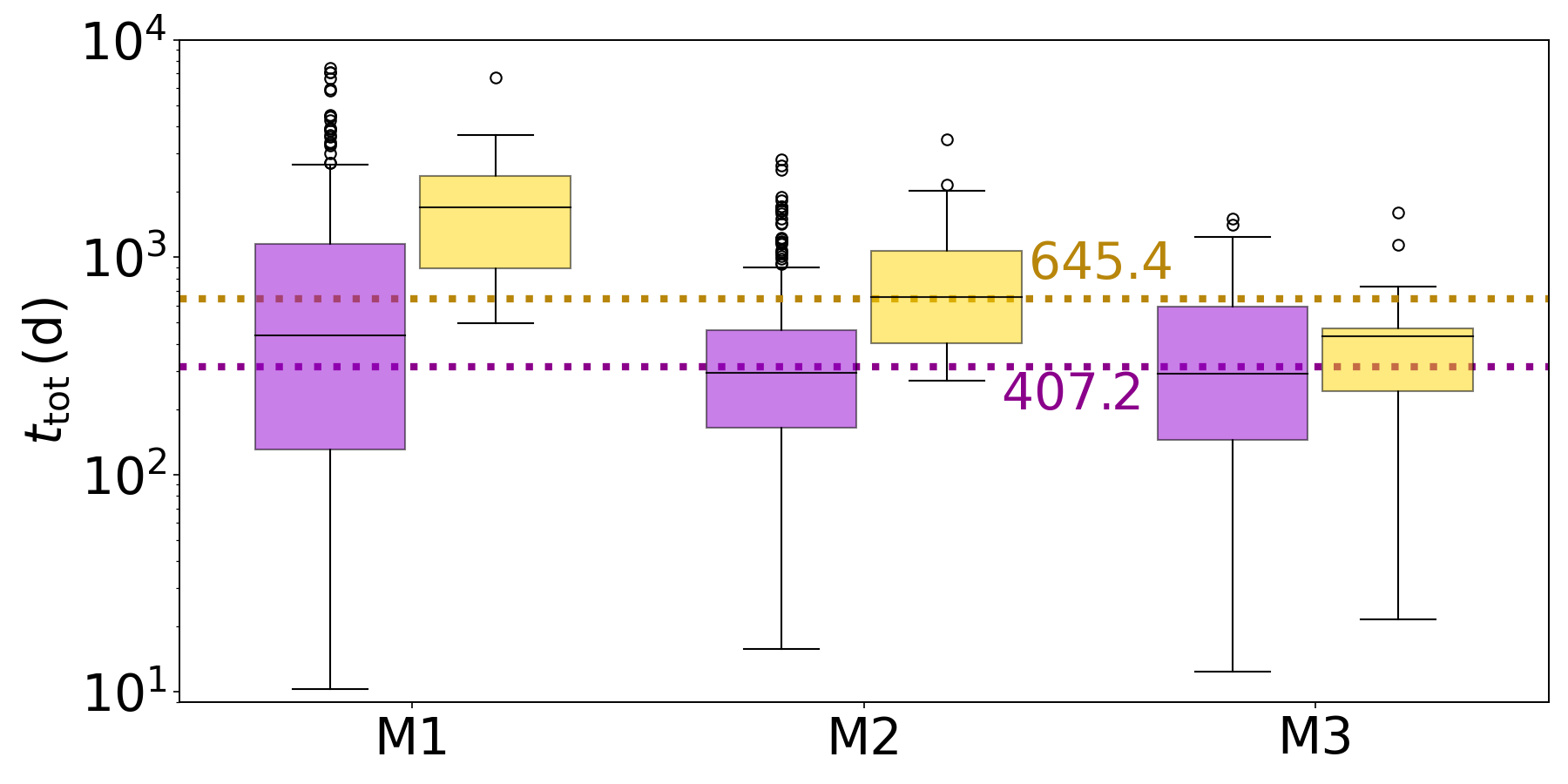} \\
    \end{tabular}
    \caption{
    Same as Fig.~\ref{fig:boxplot_group2and3}, for the photometric amplitudes (upper panel), build-up and dissipation times (second and third panels) and the
    total duration of the disk event, at the bottom part.
    }
    \label{fig:boxplot_group3}
\end{figure}

As shown in Tab.~\ref{tab:groupsmeasurements}, group 3 stars present the additional advantage of allowing for a qualitative measurements of three parameters describing an isolated disk event: its photometric amplitude and the duration of build-up and dissipation.
Figure~\ref{fig:boxplot_group3} shows normalized histograms of these three quantities for both galaxies. In addition, the total duration of the disk event ({$t_{\rm tot}=t_{\rm b}+t_{\rm d}$}) is displayed in the last panel.

The photometric amplitude, shown in the top plot, displays significant differences between the two galaxies, with those in the SMC being roughly three times as large, on average, as those in the LMC. 
Another difference lies in the mass dependency of $\Delta I$: for the LMC there are no significant differences between $M_1$ and $M_2$, but for $M_3$ this parameter shows a slight decrease.
Conversely, for the SMC, there is a systematic \emph{increase} in photometric amplitude with mass. Additionally, the spread in $\Delta I$ is significantly greater for the SMC sample, {suggesting more variability within the IQR.}

%

SMC stars systematically harbor longer disk events than their LMC counterparts. For all parameters ($t_{\rm b}$, $t_{\rm d}$, $t_{\rm tot}$) and across all mass ranges, the median values are consistently larger in the SMC.
For example, considering all mass bins, the average duration of an isolated disk event in the SMC is 645 days, compared to 407 days in the LMC (see horizontal lines in Fig.~\ref{fig:boxplot_group3}).
Interestingly, unlike the photometric amplitude, the duration parameters for the SMC exhibit a significantly smaller spread.

For the SMC, mass has an important role in determining the total duration of events.
As seen in the bottom panel of Fig.~\ref{fig:boxplot_group3},  much longer events are associated with the lowest mass. For example, the median values for $M_1$ is approximately $2000\,{\rm d}$ while for $M_3$ is approximately $400\,{\rm d}$.
For the LMC the situation is not so clear.
$M_{\rm 1}$ targets produce disks lasting for approximately $440\,{\rm d}$ and this value drops to about $300\,{\rm d}$ for $M_{\rm 2}$ and $M_{\rm 3}$ stars.

\begin{figure}
    \centering
        \includegraphics[width=.95\columnwidth]{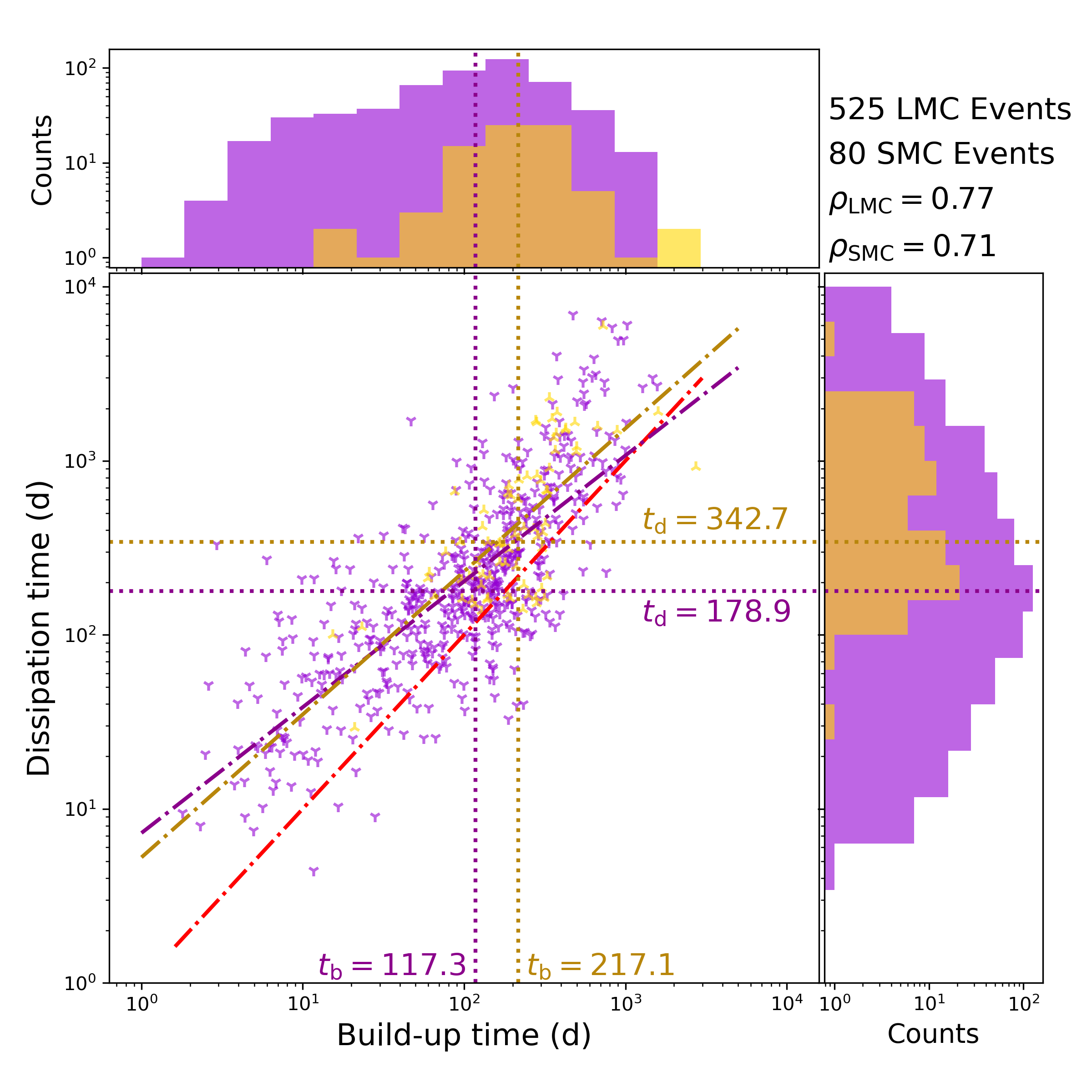} \\
    \caption{Scatter plot showing the correlation between build-up and dissipation times for the LMC (purple symbols) and SMC (yellow symbols). Shown are data extracted by fitting Eqs.~\ref{eq:rimulo1} and \ref{eq:rimulo2} to each disk event.
    The respective median values are represented by dotted lines.
    A power-law fit for each distribution is shown as dash-dot lines with the same color above, while  the identity line is shown in red.
    In the upper and right part of the figure are the histograms for build-up and dissipation times, respectively.
    In the upper right part of the plot the number of events are shown, as well as the Pearson coefficients.}
    \label{fig:scatterplot_times}
\end{figure}

Lastly, we study correlations between build-up and dissipation times.
The central panel of Fig.~\ref{fig:scatterplot_times}
shows the scatter plot for 605 disk events, 526 for LMC (purple markers) and 79 for SMC (yellow  markers).
The corresponding power-law fits are shown as dashed-dotted lines with the same color pattern. The identity line (${t_{\rm b} = t_{\rm d}}$) is indicated by the red line.
%
The corresponding histograms (not normalized) are shown in the upper and right panels, also with the same color pattern.
Additionally, in the upper right, the Pearson correlation coefficient for each galaxy is indicated, showing a moderate to strong correlation between build-up and dissipation times for both galaxies.

Figure~\ref{fig:scatterplot_times} illustrates the diversity of disk events observed in our sample. Note that only a small fraction of the stars and events are shown in this figure, as it is limited to isolated disk events. The median durations of these events, represented by the dashed horizontal (dissipation) and vertical (build-up) lines, do not accurately reflect the individual event characteristics. For example, events as short as a few days or weeks are relatively common, showcasing the complex variability patterns exhibited by Be stars.
Notably, there are a few exceptionally long disk events, some lasting over a decade. The longest isolated disk event fully captured in our sample { lasted 6708\,d, with ${t_{\rm b} = 727\,\rm d}$ and  ${t_{\rm d} = 5981\,\rm d}$. This event was observed for star S547 in the SMC.}

\section{Discussion}\label{sec:discussion}

In this section, we place the main findings from the previous section into context. 
{The classification scheme we adopted enabled us to extract the maximum amount of information from the light curves and identify homogeneous subgroups, as defined in Section~\ref{sec:data_analysis} and Table~\ref{tab:groups}. This classification approach allowed us to reliably estimate different sets of parameters from each subgroup, including stellar mass and quantitative information about disk events.
Furthermore, this study is unique in its comparison of stars from both the LMC and SMC, two galaxies with significantly different metallicities and star formation histories. This allowed us to investigate the Be phenomenon--defined here as the ability of Be stars to lose mass and form disks--across these two environments. 
By examining these differences, we were able to gain a deeper understanding of the role that mass and metallicity play in shaping the Be phenomenon.
}

\subsection{Sample overview}
\label{discussion:sample}

Groups 2 and 3 offer the possibility of estimating the central star's mass by comparing the photometric baseline levels -- representing the star's inactive state -- with theoretical photospheric models for Be stars.
The normalized histograms presented in Fig.~\ref{fig:Mass_hist} illustrate the mass distributions of the analyzed sample. 
A notable feature of these histograms is the significant difference in the number of LMC stars {(283)} compared to SMC stars {(80)}. This disparity arises from two factors.
First, the raw sample of SMC stars is inherently smaller, containing {989} stars compared to {2,144} stars in the LMC. Second, there is substantial evidence that SMC stars exhibit much higher levels of activity than their LMC counterparts (as elaborated in the next subsections). This increased activity means it is less common to observe SMC stars with well-defined photometric baselines, which is a prerequisite for inclusion in groups 2 and 3.
As a result, the sizes of groups 2 and 3 for the SMC are significantly reduced compared to those of the LMC. This bias highlights a methodological limitation: stars with higher variability are systematically underrepresented in the mass analysis. Consequently, the histograms do not fully reflect the underlying mass distribution of the entire Be star population but are instead influenced by the constraints of the selection process.

{Despite the variability bias commented earlier, the mass distribution} aligns well with estimates from the literature. These estimates were derived by combining the IMF with the Be star ratio (Be/(B+Be)). This agreement suggests that our LMC sample provides a reasonably accurate representation of the overall Be star population in the LMC galaxy. Thus, we can reasonably infer that the observed trends and distributions for the LMC sample are reflective of its broader stellar population.
In contrast, the situation for the SMC is notably different. Figure~\ref{fig:Mass_hist} highlights stark discrepancies between the mass distribution of our SMC subsample and two other SMC distributions estimated using data from the literature. Our sample shows a pronounced dearth of low-mass stars and an excess of high-mass stars compared to these other distributions. This divergence underscores that our SMC subsample is likely not representative of the SMC's broader Be star population.

Several factors could contribute to this mismatch. First, the inherent activity bias discussed earlier may preferentially exclude low-mass stars, which are more likely to exhibit subtler photometric variability or lack well-defined baselines (however, SMC stars had larger photometric amplitudes on average).
Second, the SMC is more distant than the LMC, with a distance modulus approximately 0.5 mag fainter {(Sect.~\ref{sec:raw_sample})}. This increased distance makes fainter Be stars in the SMC harder to detect compared to their LMC counterparts.

As we analyze the relationships between measured quantities and stellar mass in the sections below, it is important to keep this limitation in mind. While the LMC results can likely be generalized to the galaxy as a whole, caution must be exercised when interpreting trends for the SMC. Future efforts could aim to address this bias by increasing the SMC sample size, exploring additional observational strategies to better capture variability in low-mass stars, or employing statistical corrections to account for sample selection effects. Such approaches would enable a more robust understanding of the Be star phenomenon across both galaxies.

\subsection{Overview on Mass Loss}\label{sec:discussion_mass_loss}

The variability diagnostics explored in Sect.~\ref{sec:results_wholesample} are available for the entire sample of Be star candidates (Groups 1 to 3).
In this study, we measure, for the first time, a fundamental quantity related to the mass-loss properties of Be stars: the duty cycle (DC). Together with the number of outbursts, $N_{\rm out}$, these parameters provide critical insights into the mass-loss behavior of Be stars, including the frequency and duration of outburst events.
Before discussing the main properties derived for these two parameters, it is important to acknowledge the inherent biases present in our sample. Be stars were identified based on their variability patterns, as detailed in Sect.~\ref{sec:raw_sample}. Consequently, many stars that were excluded from the original sample due to a lack of detectable activity might still be inactive Be stars.
Therefore, the median values of DC and $N_{\rm out}$, as shown in Figs.~\ref{fig:hist_Nout} and ~\ref{fig:hist_dutycycle}, should be interpreted as upper limits rather than definitive measurements.

The overview provided by these two quantities reveals an intriguing pattern. While the distribution of the number of outbursts is relatively similar for both galaxies, with a median value of approximately 0.3 outbursts per year, the duty cycle (DC) exhibits notable differences. Stars in the SMC are significantly more active than those in the LMC. On average, SMC stars spend 60\% of their time undergoing mass loss, compared to just 44\% for LMC stars.


The disk duty cycle (DDC) is another parameter measured for the first time in this study. By using the dissipation times in the numerator of Eq.~\ref{eq:disk_duty_cycle}, the DDC provides a measure of the proportion of time a star has an observable disk.
There are two important considerations regarding this parameter. First, similar to the DC, the DDC is subject to the activity bias, making it an upper limit. 
Second, the $I$ band primarily traces emission from the inner disk. Consequently, when the emission returns to baseline levels (as observed for star LMC\_SC1\_164770 in Fig.~\ref{fig:example_groups}), this does not necessarily imply that the entire disk has dissipated.
For example, during the dissipation phases of the Galactic Be star $\omega$~CMa, studied by \citet{Ghoreyshi_2018MNRAS.479.2214G} and {\citet{Ghoreyshi2021ApJ...909..149G}}, the $V$-band excess vanished much faster than the H$\alpha$ emission. This behavior, which is also common in other Be stars, suggests that the inner disk disappears more rapidly while the outer regions persist.
Therefore, the DDC values reported here are likely underestimated, as the disks are present for longer periods than indicated by $I$-band observations alone.
Consequently, our results indicating a median DDC value close to 1 should be interpreted with caution. 
On one hand, the first bias—stemming from the exclusion of inactive Be stars—suggests that the reported DDC values are overestimated, as the sample is skewed towards stars exhibiting high levels of variability and activity. On the other hand, the second bias—related to the limitations of using $I$-band photometry to track disk dissipation—implies that the DDC values are underestimated, as they do not fully account for the persistence of outer disk regions after the inner disk emission fades.
These opposing effects create a significant uncertainty in the true value of the DDC. While the observed high median values suggest that Be stars are highly active and maintain disks for much of their observed lifetimes, the interplay of these biases prevents a definitive conclusion. Further studies, incorporating multi-wavelength observations (e.g., combining photometric and spectroscopic data) and larger, unbiased samples, are essential to refine the measurements of the DDC and better understand the longevity and variability of disks around Be stars.

The mass determination for stars in groups 2 and 3 provides valuable insights into the mass dependency of key diagnostics related to mass loss (DC and $N_{\rm out}$) and disk survival (DDC).  
For both metallicities, the results clearly indicate that mass loss and disk activity increase with stellar mass (as seen in Fig.~\ref{fig:Mass_hist}).
This trend is evident across all diagnostics, but the most pronounced dependency is observed for $N_{\rm out}$. 
Specifically, more massive stars tend to exhibit higher duty cycles, indicating they spend a larger fraction of their time actively losing mass. Similarly, the number of outbursts also increases with mass, suggesting that massive stars experience more frequent episodes of enhanced mass loss.  
This strong mass dependency aligns with Galactic studies \citep{Labadie-Bartz_2018AJ....155...53L} and recent results that indicate that the pulsational amplitude in Be stars increases with mass \citep{Labadie-Bartz2022AJ_163_226}.

\subsection{Overview on isolated disk events}
\label{sec:discussion_isolated}

The ideal conditions for studying disk event features occur in isolated events, which are not influenced by a preexisting disk, making them  easier to study and model \citep{Rimulo_2018MNRAS.476.3555R}.
An example can be found in the bottom panel of Fig.~\ref{fig:example_groups} for LMC\_SC1\_164770.
In contrast, the lightcurve at the middle panel (SMC\_SC2\_29957) begins with a significant disk whose initial conditions are difficult to determine.
%

We identified 605 such events (525 in the LMC and 80 in the SMC) in our sample.
Using Eqs.~\ref{eq:rimulo1} and \ref{eq:rimulo2}, we measured the photometric amplitude and accurately determined the build-up and dissipation durations.  
While the photometric amplitude shows no significant mass dependency, it exhibits a strong metallicity dependency. Lower metallicity is associated with significantly more vigorous disk events, as indicated by the typical values of $\Delta I$: the SMC has an average of $-0.44\,{\rm mag}$, almost three times the LMC's average of $-0.15\,{\rm mag}$.  

From Fig.~\ref{fig:modelos_angulointermediario}, we observe that larger photometric amplitudes correspond to denser disks. Even when accounting for uncertainties in the inclination classification of our sample-- where stars were broadly categorized as either pole-on or edge-on based on the CMD shape (see Sect.~\ref{sec:models_dynamic})--we can confidently conclude that SMC disks are, on average, {denser and/or longer fed than their LMC counterparts.}

A clear correlation between the build-up and dissipation times is evident in our data, despite considerable scatter.
This underscores the mass reservoir effect (MRE) described earlier: a disk that has been fed for a longer period accumulates a substantial mass reservoir in its outer regions. This reservoir sustains the disk during the dissipation phase, significantly slowing the dissipation rate.


\citet{Jian_2024A&A...682A..59J} found similar results: dissipation phases (their decay) last longer (524 days) than build-up (474 days). 
A similar approach to \citeauthor{Jian_2024A&A...682A..59J} was also used by \cite{Bernhard_2018MNRAS.479.2909B} to study visible data of Galactic Be stars, obtaining similar results, with average dissipation phase (their falling time) of $51\,{\rm d}$, while rising last 28.5\,d on average.

\subsection{Metallicity impacts in the manifestation of the Be phenomenon}\label{sec:disc_metallicity}

Our results indicate that lower metallicity stars spend more time losing mass and sustaining a disk; also, they tend to harbor longer-lived events with stronger photometric variability, while outburst rates do not show significant correlation.
This is supported by the results of \cite{Martayan_2007A&A...462..683M}, who identified that low Z stars rotate at much higher rates \citep[confirming theoretical predictions from evolutionary models, ][]{Meynet_2000A&A...361..101M,Maeder_2001A&A...373..555M}.
These higher rotation rates at low metallicity may indicate that it is easier to initiate and sustain mass loss in Be stars. This could be a significant factor in explaining the observed behavior in our sample.

Higher rotation rates at low metallicity can also help explain the larger fraction of Be stars found in the SMC compared to the LMC. However, our sample, being highly inhomogeneous, cannot be used to support this finding, which has already been reported by many authors \citep[e.g.,][]{Martayan2006A&A...452..273M,Martayan_2007A&A...462..683M,Martayan2010A&A...509A..11M}.


As mentioned in the previous section, the majority of light curves with isolated events originate from LMC stars. A possible explanation for this discrepancy lies in the stronger correlation between Be star activity and metallicity. SMC stars are more active and often exhibit frequent and overlapping disk events.
This increased activity has two notable consequences for the SMC sample:
i) A smaller number of stars with clear baselines (group 2), as the higher activity levels make it challenging to identify and classify stars that do not display ongoing or recent mass loss.
ii) A reduced number of isolated events (group 3), as the more frequent occurrences of overlapping or consecutive disk events prevent many SMC stars from undergoing distinct, isolated outbursts.

\subsection{Relevance of the stellar mass on Be star activity}\label{sec:disc_mass}


Our results indicate that all metrics of Be star activity (DC, DDC and $N_{\rm out}$) increase with stellar mass for both galaxies: early-type stars tend to display higher values of DC, DDC, and more frequent outbursts.
A correlation between the last parameter and mass was also found by \citet{Bernhard_2018MNRAS.479.2909B} based on a sample of Galactic stars.
The measured quantities in their study are larger than ours, with the mean values (which are more susceptible to outliers) of $N_{\rm out} = 0.2\pm0.4$, for late type (B7 or later), $0.7\pm1.8$, for intermediate (B4, B5 and B6) and $2.3\pm3.1\,{\rm yr^{-1}}$ for early type (earlier than B4) stars.
\citet{Labadie-Bartz_2017AJ....153..252L} also estimated this quantity, finding a typical value of $\sim 4\,{\rm yr^{-1}}$ also for a sample of Galactic stars.
It was also identified that massive stars favor the emergence of short-lived disks, which is consistent with other works that found earlier spectral types exhibit a higher degree of variability \citep[for example,][]{Cuypers1989A&AS...81..151C,Gutierrez-Soto2008CoAst.157...70G,McSwain2009ApJ...700.1216M,Chojnowski2015AJ....149....7C,Labadie-Bartz_2017AJ....153..252L}.
{It is worth noting that timescales that were more accessible for Galactic samples were faster, as a result of higher photometric precision and shorter cadence, and the absolute numbers are not easy to compare, but the trends with mass are all agreeable.
}
While our results do not show a clear correlation with $\Delta I$, \citet{Bernhard_2018MNRAS.479.2909B} report that the amplitude of Galactic outbursts increases with increasing stellar mass.


Previous studies have identified a correlation between stellar mass and the maximum density of the disk \citep[e.g.,][]{Vieira_2017MNRAS.464.3071V, Rimulo_2018MNRAS.476.3555R, Arcos2017ApJ...842...48A}. Given the relationship between disk density and the amplitude of photometric variations, a correlation between mass and $\Delta I$ was expected in our data.
The absence of such a correlation in our results could stem from the fact that $\Delta I$ is influenced by additional factors, such as the inclination angle and the viscosity parameter, which were not constrained in our analysis. Only a detailed light curve modeling, similar to the approaches employed by \citet{Rimulo_2018MNRAS.476.3555R} and \citet{Ghoreyshi_2018MNRAS.479.2214G}, could accurately determine this relationship.

\section{Conclusions}

We analyzed the largest photometric dataset of Be stars to date, consisting of 3133 light curves (2144 from the LMC and 989 from the SMC), covering nearly 20 years of observations. Together, these data account for approximately 60 millennia of observational time for the LMC and SMC galaxies combined.
This study is the first to analyze {large Be star samples} from both galaxies, enabling an investigation into the impact of metallicity on the Be phenomenon.

A novel methodology was introduced to investigate Be star activity. Leveraging the viscous decretion disk (VDD) theory, which successfully reproduces many observed features of Be stars, we ran a set of dynamical models to interpret the photometric variability in the light curves (Sect.~\ref{sec:models_dynamic}).
This approach enabled us to translate observed variability into detailed insights about Be star activity patterns. In several cases, we inferred physical parameters such as stellar mass (Sect.~\ref{sec:model_disk-less}) and estimated inclination angles (Sect.~\ref{sec:models_dynamic}), incorporating them into our analysis.
The high observational cadence of OGLE (on the order of a few days) facilitated the detection and characterization of changes in the mass-loss rate with unprecedented detail.

After a careful analysis of the entire sample to remove stars without variability and {eclipsing binaries}, we ended up with a sample of 1751 Be star candidates for which we tracked the variability behavior and estimated the orientation, by placing them into pole-on and edge-on bins.
Notably, pole-on stars constitute approximately 94\% (1639) of the entire sample.

This sample was further categorized into three distinct groups, as summarized in Fig.~\ref{fig:fluxogram_groups} (with detailed counts presented in Tab.~\ref{tab:groups}). 
Group 1 stars, which make up the majority of the sample, lack a baseline phase—defined as a segment of the light curve free from any signs of disk emission that could be used to determine the photometric brightness of the diskless state. For this group, we were able to measure variability diagnostics, such as duty cycle (DC), the disk duty cycle (DDC), and the outburst rate ($N_{\rm out}$). 
Group 2 stars possess a baseline phase, which allowed us to additionally estimate their mass. 
Group 3 stars, in addition to having a baseline phase, presented the important feature of isolated disk events, from which we measured the photometric amplitude and duration of the disk build-up and dissipation phases.

The main conclusions of this work are as follows:

\begin{itemize}
    \item Be stars in the SMC are generally more active than their LMC counterparts,
    show longer-lived disk events but display slightly smaller outburst rates.
    
    \item High-mass stars consistently exhibit greater activity than low-mass stars across all variability metrics analyzed.
    The strongest trend with mass is in the outburst occurrence rates, with high-mass stars exhibiting on average an outburst rate 3 times higher than the lower mass stars. At the same time, the duration of outbursts systematically decreases with mass (especially for the SMC).

    \item The DC is a fundamental quantity that measures the fraction of time the star is actively losing mass and feeding a disk.
    The DCs measured for both galaxies are likely upper limits, because inactive Be stars were removed from the analyzed sample. The same is true for the DDC and $N_{\rm out}$.
    There is significant diversity in the DC for both samples, ranging from 0 to 1. The DC distribution for the LMC is broad, with a median value of 0.44, whereas the distribution for the SMC is clearly more concentrated at higher values, with a median of 0.60. It is worth noting that inactive Be stars (not analyzed) would contribute only to the very first bin of the respective histograms and would not otherwise alter the overall shape of the distributions.

    \item For the analyzed sample and both galaxies, the median DDC is close to one, indicating that the disk is present for most of the observational period. 
    The median value may reflect a selection bias, as our sample was specifically chosen for exhibiting disk activity at some point in the light curve, and is likely an upper limit.
    However, as noted above, adding the inactive Be stars to the sample would not change the shape of the DDC distributions.


    \item For a small subsample (group 3), important quantities of isolated disk events--such as photometric amplitude and the durations of build-up and dissipation--were measured.
    Two significant results were obtained: SMC stars exhibit much larger photometric amplitudes and, consequently, likely possess denser disks than their LMC counterparts. Additionally, the total durations of their disk events are systematically longer.

    \item The duration of build-up and dissipation of group 3 are strongly correlated, as previously reported in the literature and predicted by the VDD theory. This confirms the mass reservoir effect described by \citet{Rimulo_2018MNRAS.476.3555R}. 
    For the SMC, $t_{\rm d} / t_{\rm b} \approx 1.6$, and for the LMC the ratio is 1.5, on average.
\end{itemize}

Two natural follow-up studies are planned. The first involves applying the same methodology to a sample of Galactic Be stars, allowing for direct comparisons across environments with differing metallicities. The second focuses on detailed modeling of the hundreds of isolated disk events identified in this study. This effort will yield robust estimates of disk densities in the SMC and LMC.
Together, these studies will provide deeper insights into how metallicity influences the mass-loss mechanisms and disk formation processes in Be stars.





\begin{acknowledgments}
A.\,L.\,F. acknowledges support from CAPES (grant 88882.332925/2019-01) and FAPESP (grant 2023/06539-0).
A.\,C.\,C. acknowledges support from CNPq (grant 314545/2023-9) and FAPESP (grants 2018/04055-8 and 2019/13354-1). 
T.\,H.\,A acknowledges support from FAPESP (grant 2021/01891-2).
%
This work made use of the computing facilities of the Centro de Processamento de Dados do IAG/USP (CPD-IAG), whose purchase was made possible by the Brazilian agency FAPESP (grants 2019/25950-8,  2017/24954-4 and 2009/54006-4).
\end{acknowledgments}




\appendix

\section{Detection threshold}\label{sec:app_observationalthreshold}

\begin{figure}
    \centering
    \begin{tabular}{cc}
    \includegraphics[width=.45\textwidth]{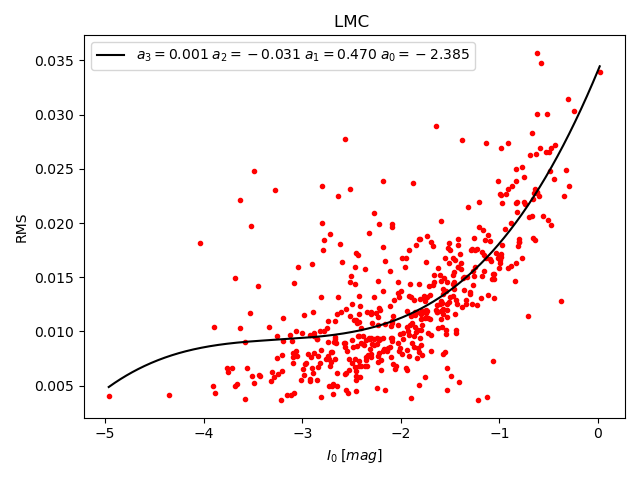} & \includegraphics[width=.45\textwidth]{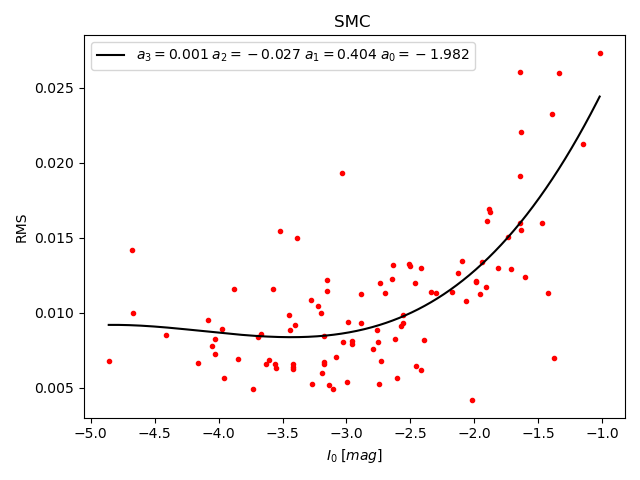} \\
    \end{tabular}
    \caption{Root mean square measure from baseline phases of group 2 and 3 stars as a function of their respective median absolute brightness value. Left: LMC. Right: SMC. The black line shows the third-degree polynomial fit, with respective parameters shown in the upper part of each panel.
 }
    \label{fig:threshold}
\end{figure}

Based on well-sampled baseline phases (containing more than 5 data points each) of light curves of groups 2 and 3, the root mean square, RMS, was estimated as a function of the median value of corresponding brightness, as shown in the Fig.~\ref{fig:threshold}.
For both galaxies, the upper limit of the RMS is about 0.03.
Based on this, a conservative estimate of the detection threshold of disk events of $0.06\,\mathrm{mag}$ was adopted. Only events with amplitudes larger than this threshold were considered in the analysis.

\section{$V-I$ CMD for low density models.}\label{sec:app_CMDlowdensity}

\begin{figure}
    \centering
    \begin{tabular}{c c}
        \includegraphics[width=0.45\linewidth]{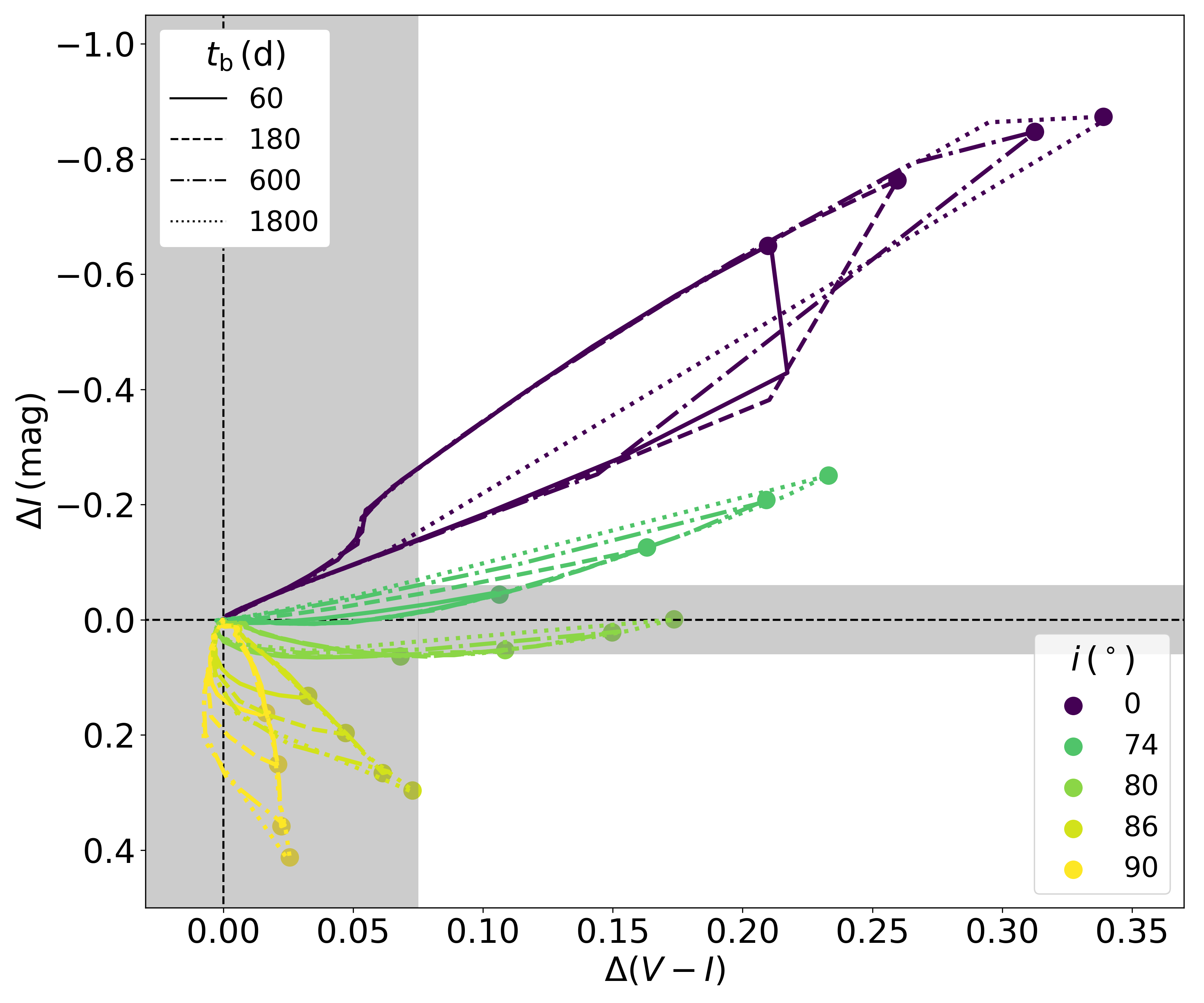} &
        \includegraphics[width=0.45\linewidth]{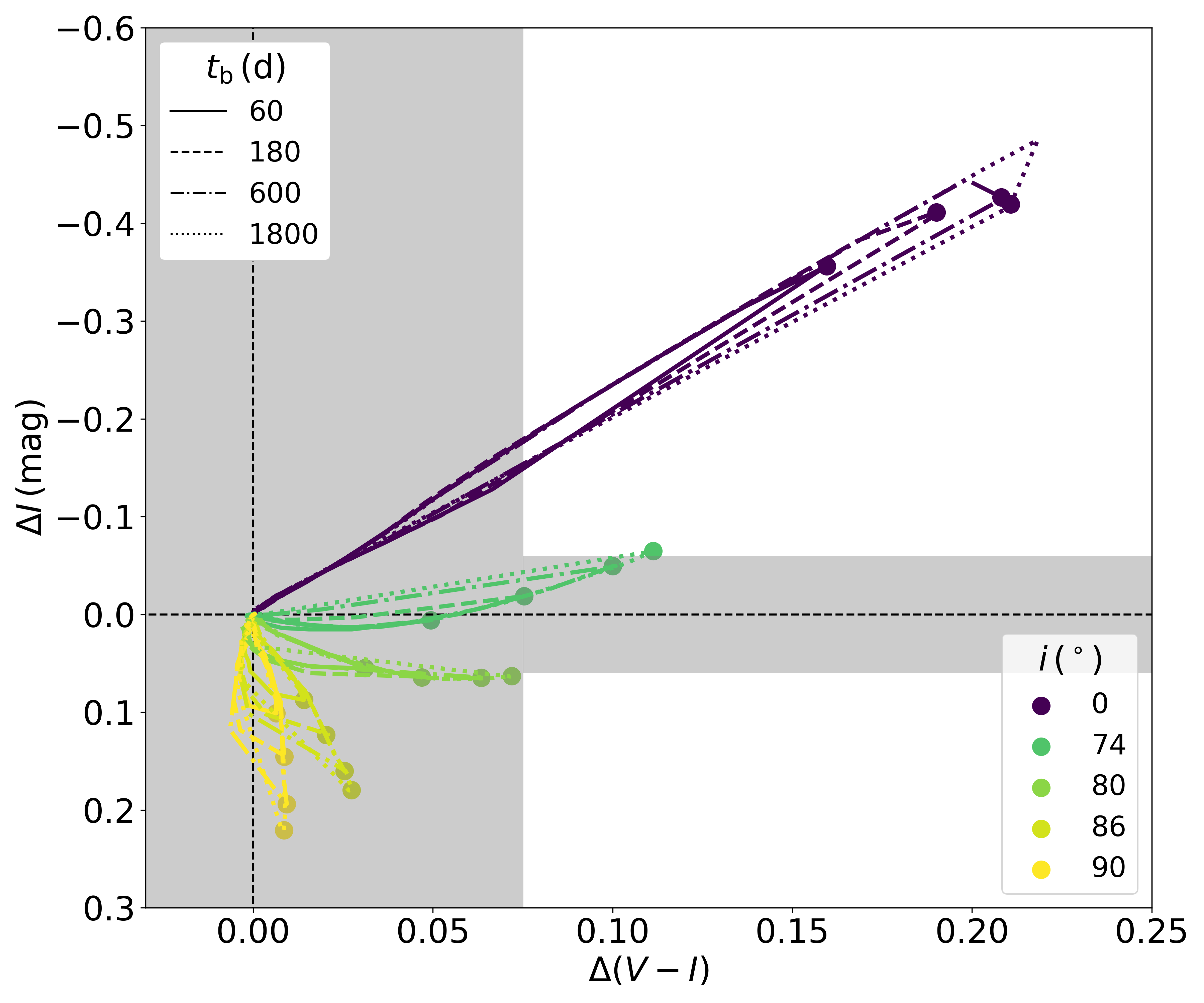}\\
        \includegraphics[width=0.45\linewidth]{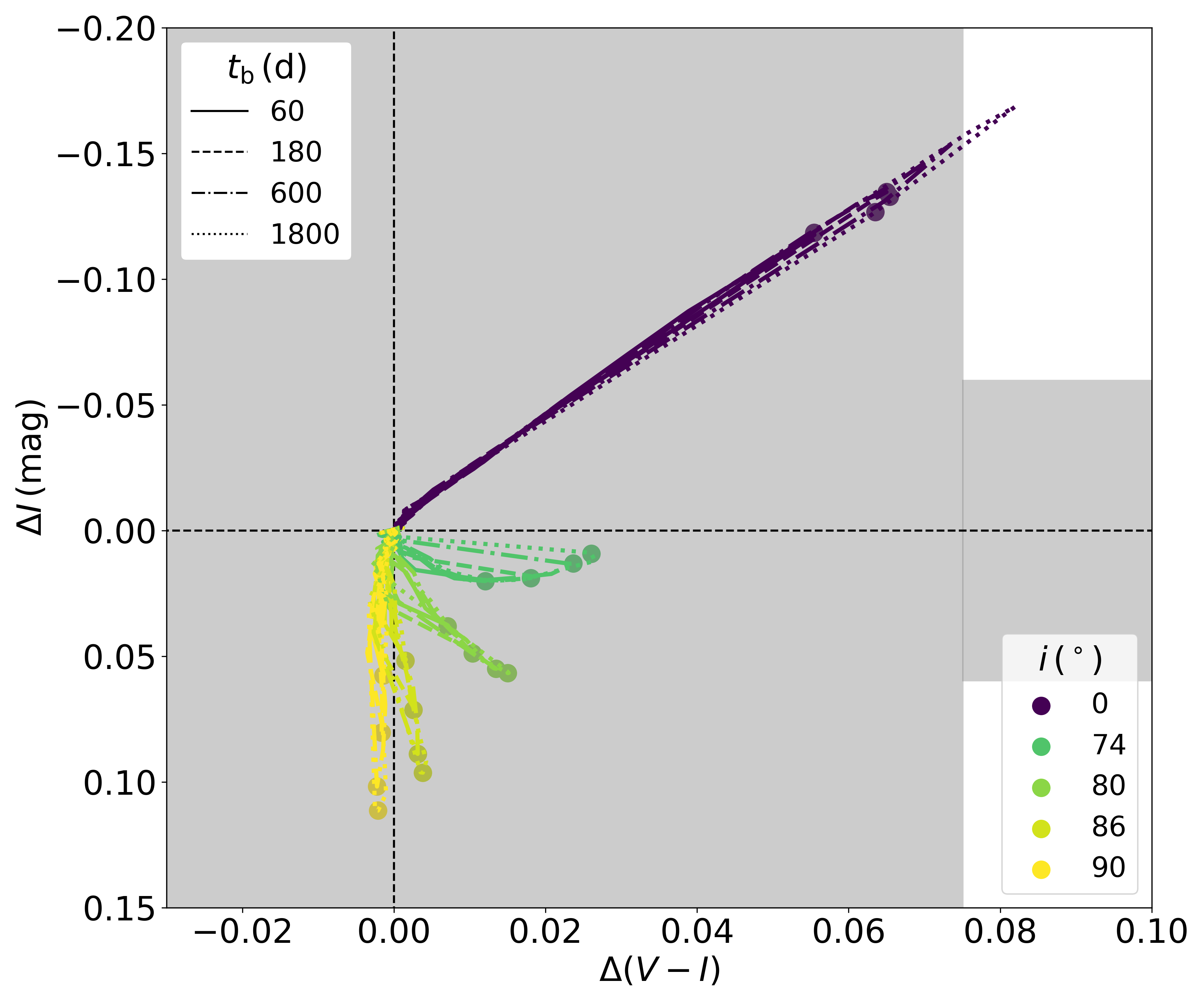} & 
        \includegraphics[width=0.45\linewidth]{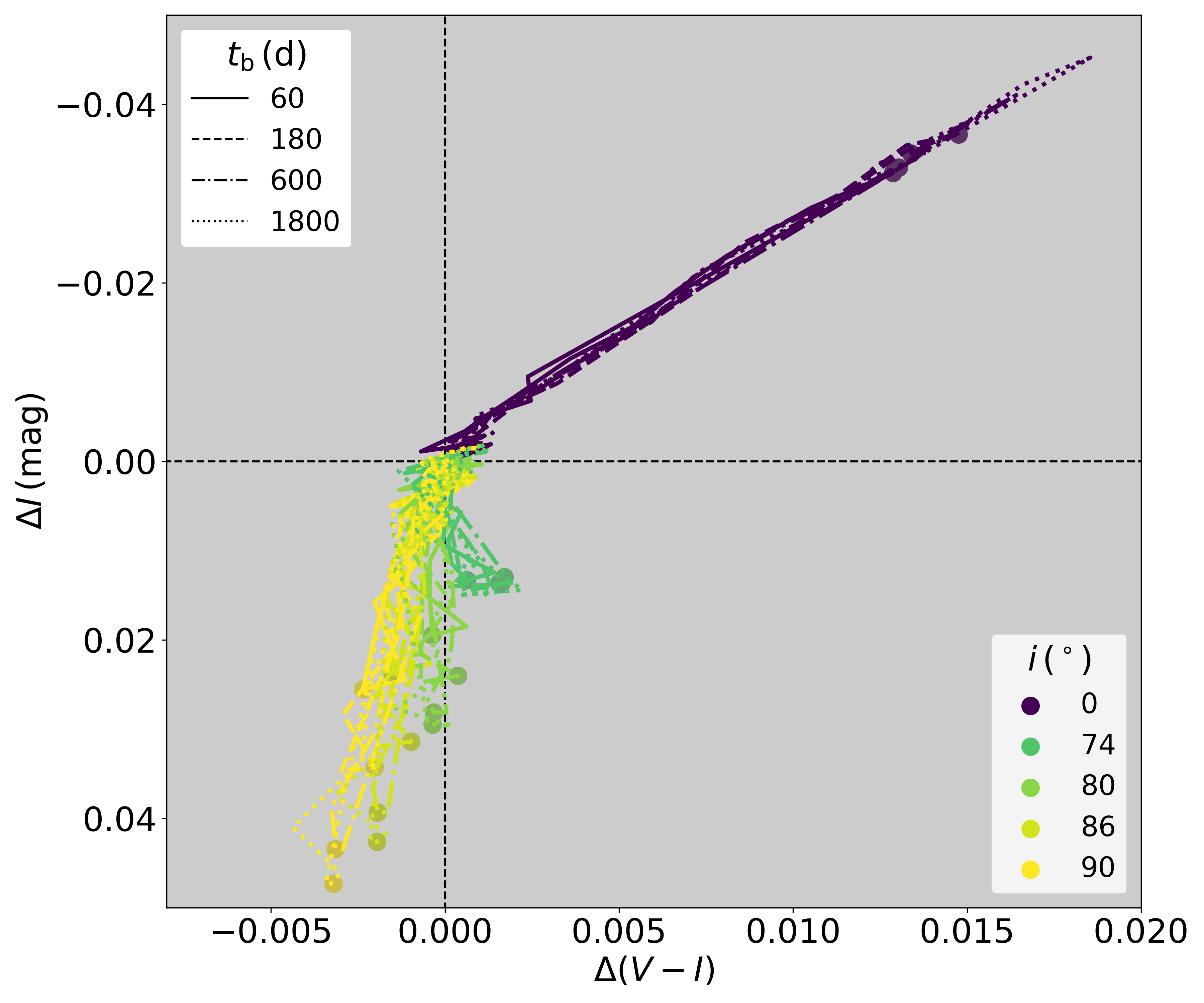}\\
   \end{tabular}
    \caption{Same as Fig.~\ref{fig:modelos_CMD_tempos}, for $\Sigma_0 = 1.67$ (top left), $0.69$ (top right), $0.29$ (bottom left) and $0.12\,{\rm g\,cm^{-2}}$ (bottom right).}
    \label{fig:app_cmd_densities}
\end{figure}

The respective $V-I$ CMD for densities $\Sigma_0 = 1.67$, $0.69$, $0.29$ and $0.12\,{\rm g\,cm^{-2}}$ are shown in Fig.~\ref{fig:app_cmd_densities}. 

\bibliography{artigo_submited}{}
\bibliographystyle{aasjournal}

\end{document}